\RequirePackage{amsmath}
\documentclass{llncs}
\usepackage{multirow}
\usepackage{booktabs}
\usepackage{graphicx} 
\usepackage{bmpsize}
\usepackage{url}
\usepackage{placeins}
\usepackage{mathtools}
\usepackage{enumitem}
\setlist{nolistsep,leftmargin=*}
\usepackage[font=small,skip=0pt]{caption}
\usepackage{epstopdf}
\usepackage{setspace}
\usepackage{subcaption}
\usepackage{wrapfig}
\usepackage{mathtools}

\DeclarePairedDelimiter\floor{\lfloor}{\rfloor}
\usepackage{lscape}
\DeclareGraphicsExtensions{.eps}

\everypar{\looseness=-1}
\usepackage[open,openlevel=1]{bookmark}
\usepackage[font=scriptsize,labelfont=bf,skip=3pt]{caption}

\usepackage[dvipsnames,table]{xcolor}
\usepackage{hyperref}
\hypersetup{
	colorlinks   = true,    
	urlcolor     = BrickRed,    
	linkcolor    = BrickRed,    
	citecolor    =  BrickRed,      
	bookmarksopen=true
}

\usepackage{xcolor}

\begin{document} 
\title{Link Classification and Tie Strength Ranking in Online Social Networks with Exogenous Interaction Networks}
\author{Mohammed Abufouda and Katharina A. Zweig}
\institute{University of Kaiserslautern\\ Computer Science Department\\ Gottlieb-Daimler-Str. 48\\67663 Kaiserslautern, Germany\\
\email{\{abufouda, zweig\}@cs.uni-kl.de}
}
\maketitle
\begin{abstract}
Online social networks (OSNs) have become the main medium for connecting people, sharing knowledge and information, and for communication. The social connections between people using these OSNs are formed as virtual links (e.g., friendship and following connections) that connect people. These links are the heart of today's OSNs as they facilitate all of the activities that the members of a social network can do.
However, many of these networks suffer from noisy links, i.e., links that do not reflect a real relationship or links that have a low intensity, that change the structure of the network and prevent accurate analysis of these networks. Hence, a process for assessing and ranking the links in a social network is crucial in order to sustain a healthy and real network. Here, we define link assessment as the process of identifying noisy and non-noisy links in a network. In this paper\footnote{The work in this paper is based on and is an extension of our previous work~\cite{Abufouda2015}.}, we address the problem of link assessment and link ranking in social networks using external interaction networks. In addition to a friendship social network, additional exogenous interaction networks are utilized to make the assessment process more meaningful.
We employed machine learning classifiers for assessing and ranking the links in the social network of interest using the data from exogenous interaction networks. The method was tested with two different datasets, each containing the social network of interest, with the ground truth, along with the exogenous interaction networks. The results show that it is possible to effectively assess the links of a social network using only the structure of a single network of the exogenous interaction networks, and also using the structure of the whole set of exogenous interaction networks. The experiments showed that some classifiers do better than others regarding both link classification and link ranking. The reasons behind that as well as our recommendation about which classifiers to use are presented.
\end{abstract}
 
\section{Introduction}
\label{introduction}
\begin{figure}[h]
    \centering
    \begin{subfigure}[b]{0.3\textwidth}
        \includegraphics[width=\textwidth]{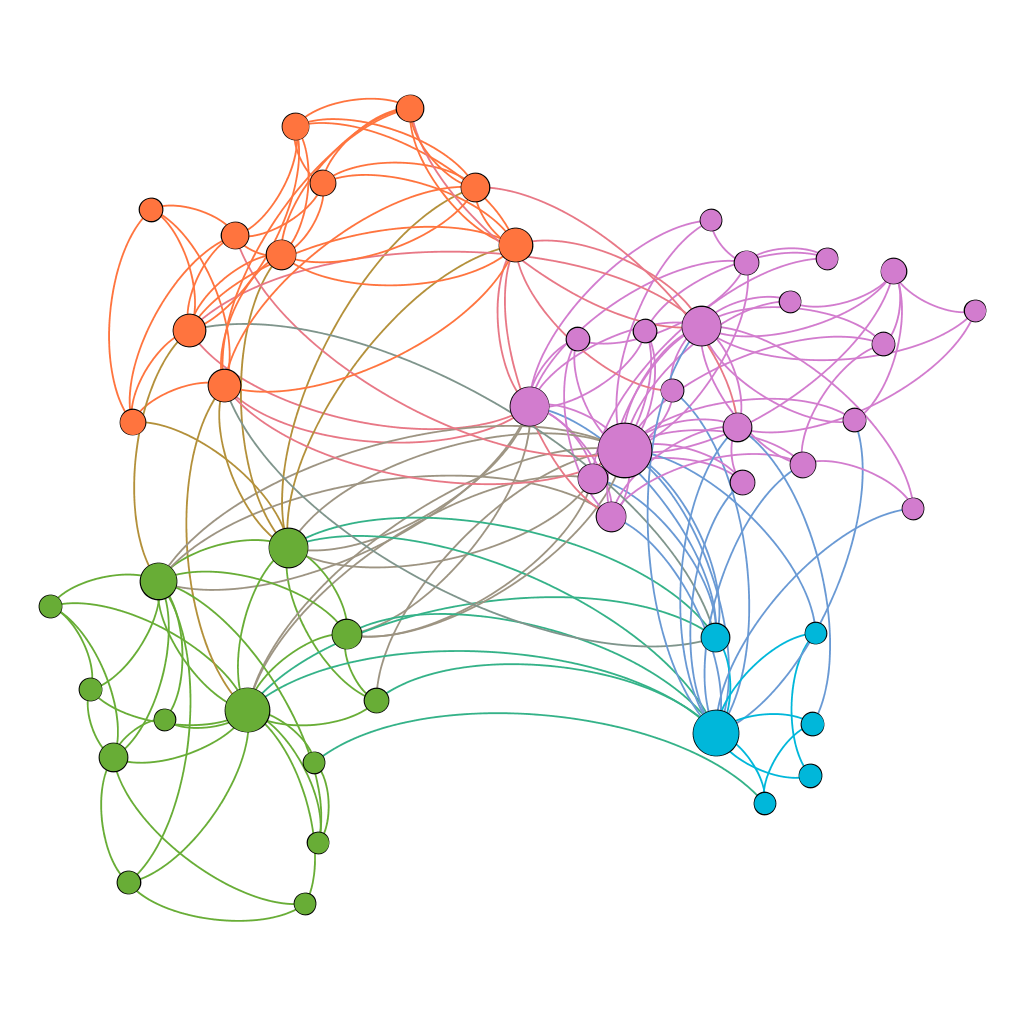}
        \caption{ {\tiny $G_1$: The Work network}}
        \label{fig:g1}
    \end{subfigure}
    ~ 
    \begin{subfigure}[b]{0.3\textwidth}
        \includegraphics[width=\textwidth]{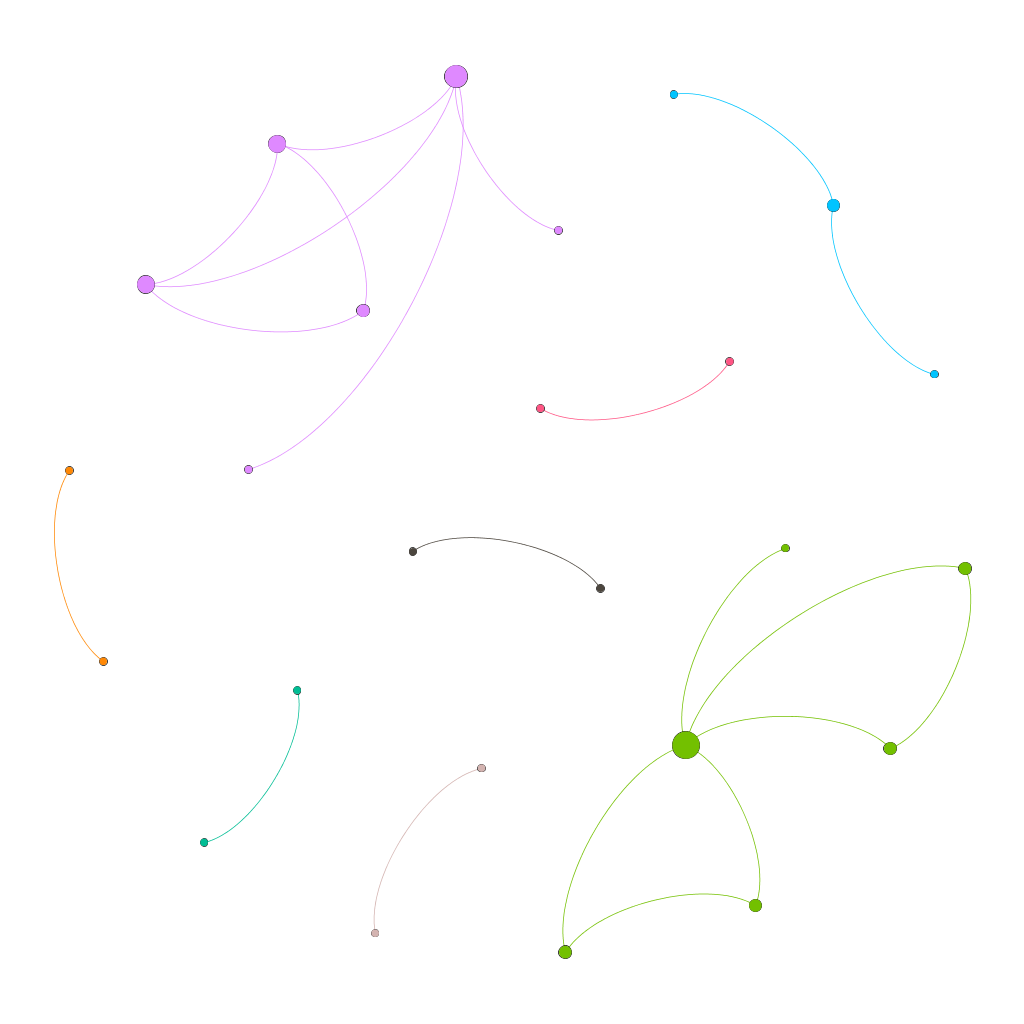}
        \caption{ {\tiny $G_2$: The Co-author network}}
        \label{fig:g2}
    \end{subfigure}
    ~ 
    \begin{subfigure}[b]{0.3\textwidth}
        \includegraphics[width=\textwidth]{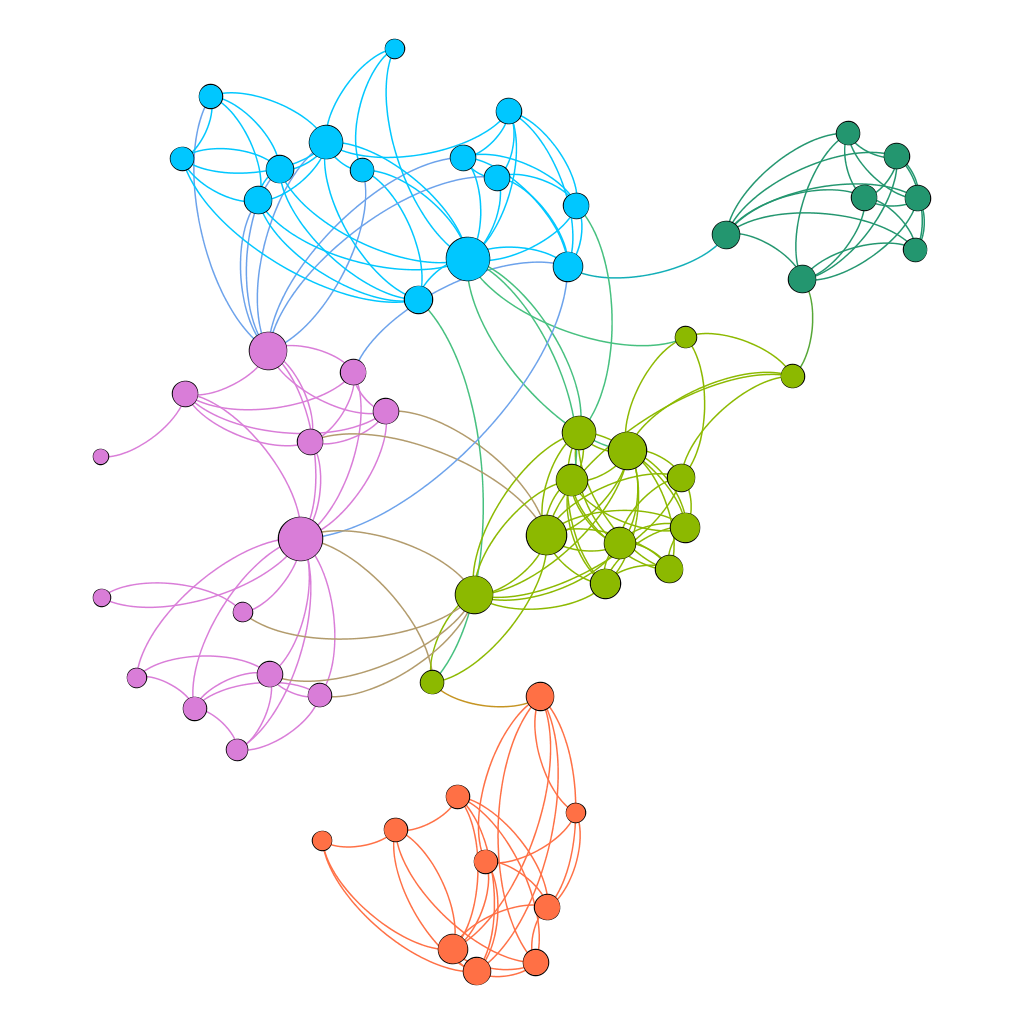}
        \caption{ {\tiny $G_3$: The Lunch network}}
        \label{fig:g3}
    \end{subfigure}
    \begin{subfigure}[b]{0.3\textwidth}
        \includegraphics[width=\textwidth]{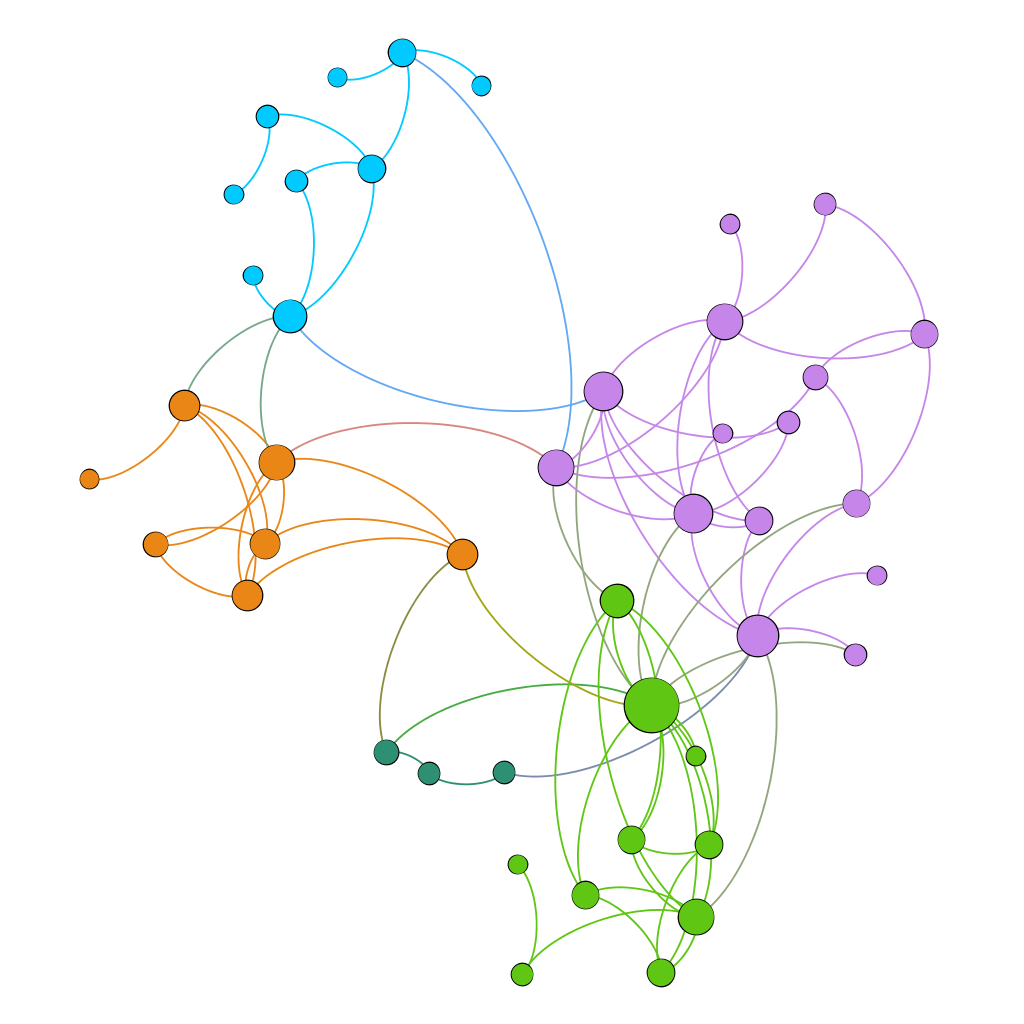}
        \caption{ {\tiny $G_4$: The Leisure network}}
        \label{fig:g4}
    \end{subfigure}
    \begin{subfigure}[b]{0.3\textwidth}
        \includegraphics[width=\textwidth]{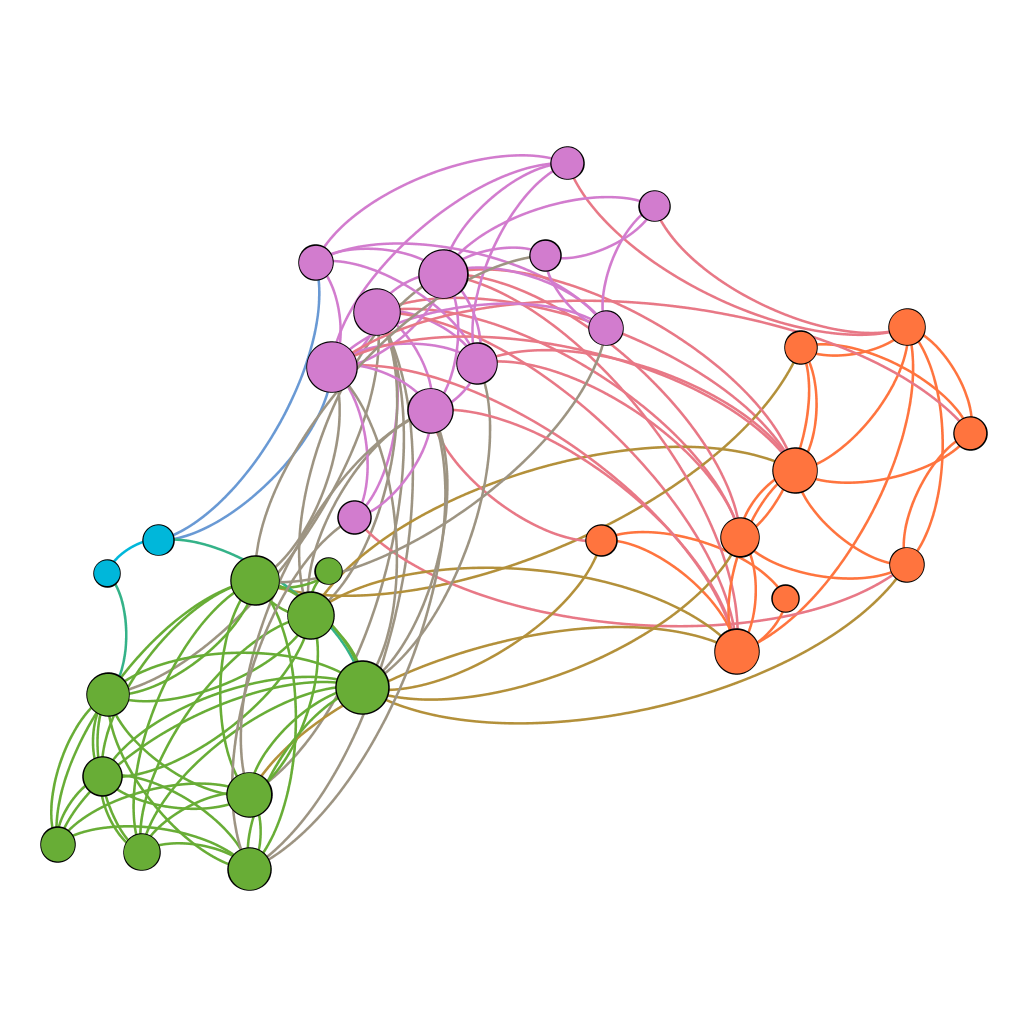}
        \caption{ {\tiny Facebook}}
        \label{fig:facebook}
    \end{subfigure}
    \begin{subfigure}[b]{0.3\textwidth}
        \includegraphics[width=\textwidth]{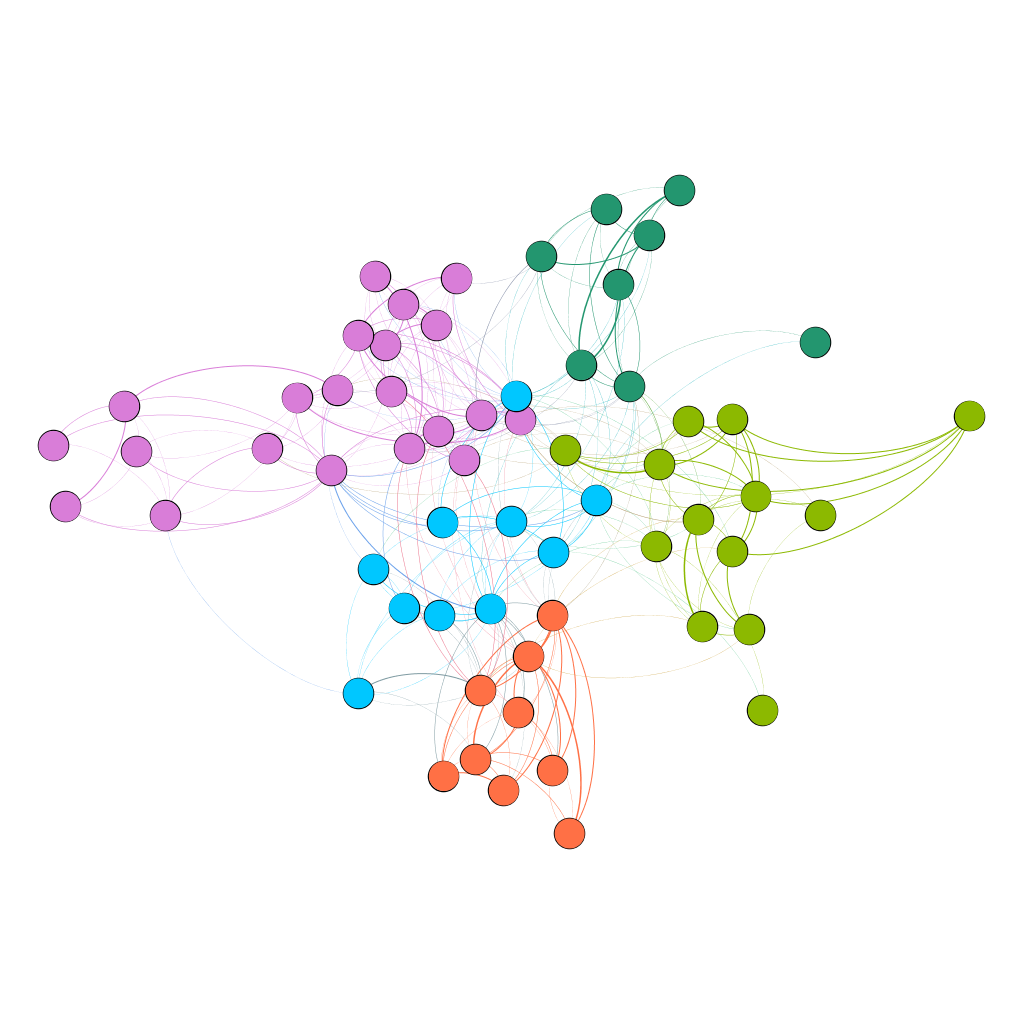}
        \caption{{\tiny  Aggregated version}}
        \label{fig:aggregated}
    \end{subfigure}
    \caption{{ (Color online) Visualization of the Research Group dataset~\cite{raey2013} networks. The visualization was done using Gephi's Yifan Hu~\cite{ICWSM09154} visualization algorithm with a few manual edits. The size of the nodes is directly proportional to their degrees and the color of the node is based on the community to which a node belongs~\cite{blondel2008}. The aggregated version in Figure~\ref{fig:aggregated} was obtained by aggregating all of the associated interaction networks into one network: $\bigcup\limits_{i=1}^{4} G_i$, where duplicated edges were removed.}}
     \label{fig:networks_RG}
\end{figure}

\begin{figure}[h]
  \centering
  \includegraphics[scale=0.4]{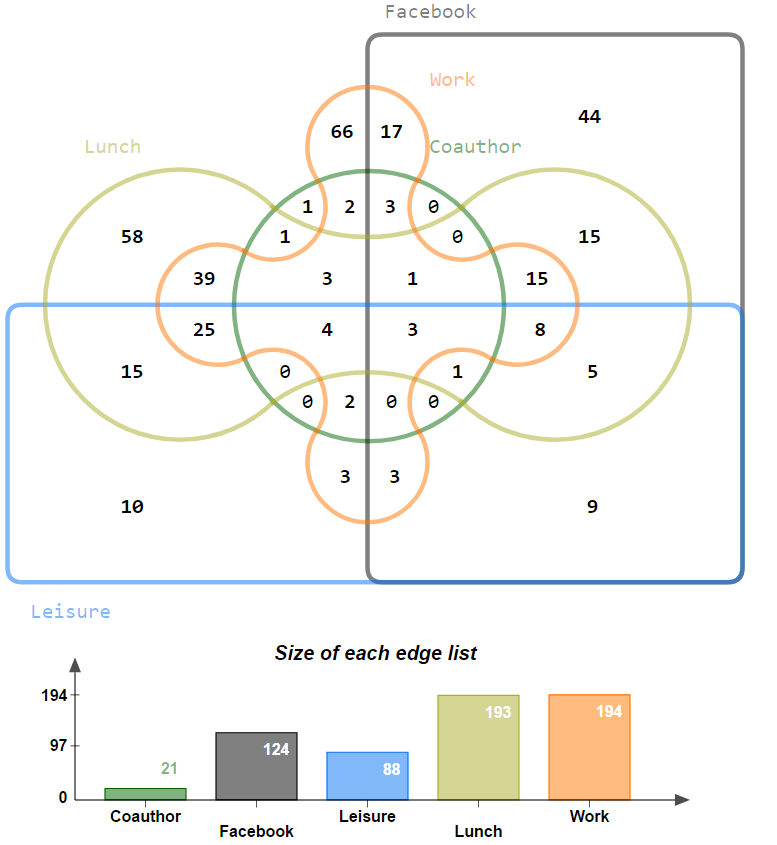}
  \caption {(Color online) The Venn diagram~\cite{bardou2014} for edge overlapping between the Facebook social network $SN$ and the other associated interaction networks $\mathcal{G}$ for the Research Group dataset.}
  \label{fig:venn}
\end{figure}

Online social networks (OSNs) have become a vital part of modern life, facilitating the way people get news, communicate with each other, and acquire knowledge and education. Like many other complex networks, online social networks contain noise, i.e., links that do not reflect a real relationship or links that have a low intensity. These noisy links, especially false-positives, change the real structure of the network and decrease the quality of the network. Accordingly, having a network with a lot of noise impedes accurate analysis of these networks~\cite{Zweig2014}. In biology, for example, researchers often base their analysis of protein-protein interaction networks on so-called high-throughput data. This process is highly erroneous, generating up to $50\%$ false-positives and $50\%$ false-negatives~\cite{Deane2002} and thus introducing noisy links into the constructed protein-protein interaction networks. As a result, assessing how real a link is in these networks is inevitable in order to get a high quality representation of the studied system. Therefore, getting accurate analysis results is hard to attain without an assessment process. Based on that, many researchers have started assessing the quality of these biological networks~\cite{Goldberg2003,chen2004} by assessing the links of these networks.
In social networks, the situation is quite similar, as many online social networks experience such noisy relationships. A friend on Facebook, a follower on Twitter, or a connection on LinkedIn does not necessarily represent a real-life friend, a real person, and a contact from your professional work, respectively.
A possible reason for the noisy relationships in these OSNs is the low cost of forming a link on online social network platforms, which results in a large number of connections for a member. Another reason for the existence of noisy relationships is the automatic sending of invitations when a member first registers on one of the social network platforms; these invitations may contribute to connecting you with persons you really do not know in real life but whom you have contacted once for any reason. Another example is the follow relationships in the Twitter social network, where it is easy to be followed by a fake account or by a real account whose owner seeks a possible follow back to get more connections.\\
In this work, we aim at assessing the relationships within a friendship social network ($SN$) based on the structure of networks related to the friendship social network of interest $SN$. These networks are called \textit{Exogenous Interaction Networks: $\mathcal{G} = \{G_1,G_2,\cdots,G_n\}$}. We have shown in previous work~\cite{abufouda2014} that exogenous interaction networks influence the tie formation process in the friendship social network; thus, using information from these networks helps to assess the links in the friendship social network. Looking merely at one individual network, in this case the $SN$, is a rather simplistic abstraction of social interaction, which is not sufficient for understanding its dynamics~\cite{Boccaletti2014}. Thus, utilizing the interaction networks that affect the structure of the social network is our concern in this work.\\
To better understand the concept of link assessment using associated interaction networks, let us consider a research center environment real data set that we will use later in the experiments. Figure~\ref{fig:networks_RG} depicts a visualization of the networks, where the members can socialize online using the Facebook social network $SN$, which is chosen as the social network of interest to be assessed later. In addition to the Facebook friendship network, the members of the research group have different interactions that affect the structure of their \textit{Facebook} friendship network $SN$. These exogenous interaction networks $\mathcal{G}$ include:
\begin{itemize}
\item Work $G_1$: Where a link exists between two members if they work/ed in the same department.
\item Co-author $G_2$: Where a link exists between two members if they have co-authored a publication.
\item Lunch $G_3$: Where a link exists between two members if they had lunch together at least one time.
\item Leisure $G_4$: Where a link exists between two members if they have participated in the same leisure activity at least one time.
\end{itemize}
The interactions in the exogenous networks affect the structure of the social network $SN$ as the link formation process within any social network is not only driven by its structure, i.e., internal homophily~\cite{McPherson2001}, but is also influenced extremely by external factors (exogenous interaction networks $\mathcal{G}$)~\cite{abufouda2014}. For example, it is highly probable that any two persons who have had lunch together and/or have spent some leisure time together will be friends in the $SN$. However, if there is a friendship link between two members \textit{A} and \textit{B} in the $SN$ and there is no link between \textit{A} and \textit{B} in any of the networks in $\mathcal{G}$, then this relationship might be a noisy one, or it may be a very low strength link that does not qualify as a real friendship relation. In Figure~\ref{fig:venn}, the links that exist in the social network $SN$ and also exist in any other network $G_i \in \mathcal{G}$ are presumably valid links, as the links in these interaction networks affect the link formation in the social network $SN$. On the other hand, there are $44$ links that are not in any $G_i \in \mathcal{G}$, which leads to the question: \textit{Are these edges noise?} In fact, these $44$ links are most likely noisy links. However, it is hard to capture all of the possible relationships between the actors of this dataset in real life. For example, one link of the $44$ might be between two researchers who are living in the same building or are members of the same political party, which is data that we do not have or that is hard to collect. Thus, these links are potential noise or relationships with very low intensity.\\
\\The reminder of this paper is structured as follows. Section~\ref{sec:relatedwork} presents the related work. Section~\ref{sec:definitions} contains the definitions and notations used in the paper. Section~\ref{sec:theproposedmethod} provides details about the method developed in this work, and Section~\ref{sec:groundtruth} defines the ground truth data set and also explains the experiment setup. Section~\ref{sec:datasets} gives details about the datasets and the evaluation metrics used to evaluate the proposed method. The paper ends with a presentation of the results in Section~\ref{sec:Results}, and the Summary in Section~\ref{sec:summary}.

\section{Related work}
\label{sec:relatedwork}
This work can be seen from two different perspectives. The first is \textit{Link prediction} using external information, and second is \textit{Tie strength} ranking. In this section, we will provide the related works based on these two categories and highlight how our work is different.\\
\textit{First}, the work in this paper is related to the link prediction problem using external information that is associated with the social networks. Hereafter, we provide the related work.\\ The problem of link prediction was initially defined in the seminal work of Liben-Nowell and Kleinberg~\cite{liben2007} followed by a plethora of research in the area of link prediction. Surveys and literature reviews such as~\cite{al2011,lu2011,wang2015,martinez2016} provided an overview of the methods used in link prediction. The most relevant work to our work is link prediction using a social network plus additional information. Wang and Sukthankar~\cite{wang2014} provided a link prediction model for predicting the collaboration of the researchers of the DBLP using different types of relations. Yang et al.~\cite{yang2012link} and Negi and Chaudhury~\cite{Negi2016} provided link prediction models for multi-relational networks, where the edges have different types of interactions. Similarly, Davis~\cite{Davis2011} provided a link prediction of Youtube following relationships using different types of interactions that were captured on Youtube such as sharing videos and sharing subscriptions. A similar work was done by Horvat et al.~\cite{zweig2012} for inferring the structure of a social network using the structures of other social networks. A recent work by Lakshmi and Bhavani ~\cite{JayaLakshmi2017} incorporated temporal data to the multi-relational dataset to provide an effective link prediction.\\
The contribution in this paper is different from the aforementioned research as follows. Our method considers not only online activity of the members as social relationships, but also some other offline interactions or interactions that are platform independent, i.e., interactions that took place outside the social network platform. This, gives more insights about the motives behind tie formations in online social networks. Additionally, to the best of our knowledge, the link assessment problem has not been addressed before in the context of social networks as complex systems. Here in this work, we provide a definition of the link assessment problem and a method to quantify the noisy links in them.\\
\textit{Second}, the work in this paper is related to the tie strength ranking research. Hereafter, we provide the related work.\\
A recent study has shown that at least $63\%$ of Facebook users have unfriended at least one friend for different reasons~\cite{Sibona2014}. According to Sibona~\cite{Sibona2014}, the reasons for unfriending include frequent or useless posts, political and religious polarization, inappropriate posts, and others. These reasons behind the deletion of a friendship connection mean that social networks suffer from noisy relationships that need to be eliminated in order to keep only the desired friends. Accordingly, it is obvious that online social networks contain many false-positive links that push the members to use the unfriend/unfollow feature or, as a less extreme reaction, categorize unwanted connections as restricted members.
Thus, a member of an online social network can easily connect to another member based on strong motivation, like being a real-life friend or participating in the same political party, or based on weak motivation like being a friend of someone they know. This variation in the type of friendship links in social networks has led some researchers to quantify the relationship's strength~\cite{Gilbert2009,Xiang2010,zhao2012,Gupte2012,Gilbert2012,Jason2013} within social networks. Pappalardo et al.~\cite{Pappalardo2012} proposed a multidimensional model to capture the strength of the ties in social networks of the same actors. Another related work was done by Xie et al.~\cite{Xie2012}, where the authors studied Twitter's users to identify real friends. Also, Spitz et al.~\cite{Spitz2016} assessed the low-intensity relationships in complex bipartite networks using node-based similarity measures. Pratima and Kaushal~\cite{Pratima2012} provided a prediction model for predicting tie strength between any two connected users of OSNs as an alternative to the binary classification of being a friend or not. 
Kumar et al.~\cite{Kumar2016} studied the weight prediction, as form of tie strength, in signed networks.
Some researchers were interested only in quantifying strong ties. Jones et al.~\cite{Jason2013} studied the interactions among the users of Facebook to identify the strong ties in the network. A similar recent work by Rotabi et al.~\cite{Rotabi2017} employed network motifs to detect strong ties in social networks.
Some applications of tie strength have been applied in different domains. Wang~\cite{Wang2016} et al. provided a social recommendation system based on tie strength prediction. McGee et al.~\cite{McGee2013} predicted the location of the users using tie strength of the members of Twitter. \\
The contribution in this paper is different from the aforementioned research as follows. Our methods employed additional information in modeling the tie strength other than the social network. The additional information not only improved the tie strength ranking, but also gives insights regarding what are the external factors that affect the interactions in the social networks. Moreover, we validated the method for link assessment using datasets that include a ground truth which is used also as a gold standard in validating the results of tie ranking among the members of the OSCs.

\section{Definitions}
\label{sec:definitions}
An undirected network $G=(V,E)$ is a tuple that is composed of two sets $V$ and $E$, where $V$ is the set of nodes and $E$ is the set of edges such that an undirected edge $e$ is defined as $e=\{u,v\} \in E$ where $u,v \in V$. For a directed network $\overrightarrow{G}=(V,\overrightarrow{E})$, a directed edge $\overrightarrow{e}$ is defined as $\overrightarrow{e}=(u,v)$ where the node $u$ is the \textit{source} and the node $v$ is the \textit{target}. For undirected networks, the degree of a node $w$ is defined as the number of nodes that are connected to it, while for directed networks the \textit{in-degree} and the \textit{out-degree} are defined as the number of edges in the network where the node $w$ is the target and the source node, respectively.

\section{The proposed method}
\label{sec:theproposedmethod}
This section presents the details of the proposed method. It starts with a general description of the designed framework followed by a detailed information about the feature engineering.
\subsection{Framework description}
The aim of this work is to assess and to rank the links in a social network $SN$ using the exogenous interaction networks of the same members of the $SN$. The proposed method benefits from the structure of these networks in order to infer with the help of a supervised machine learning classifier whether a link in the $SN$ is a true-positive or a false-positive. The idea of the proposed framework is to convert the link assessment problem into a machine learning classification problem. In the following, a description of the framework will be provided.
\subsection{Feature data model (FDM)}
\label{fdm}
The feature data model is a model that represents a network structure using topological edge-proximity features. More formally, $\forall \ v,w \in V(G_i) $ where $v \neq w$ and $G_i \in \mathcal{G}$, the feature value $F_j(v,w)$ is calculated such that $F_j \in \mathcal{F}$, and $\mathcal{F}$ is the set of features that will be described in the following section.
\subsubsection{Edge proximity features:  }
$\ $The edge proximity features are based on the following measures:
\begin{itemize}
\item \textit{The number of Common Neighbors} ($\mathcal{CN}$): For any node $z$ in a network $G$, the neighbors of $z$, $\Lambda(z)$, is the set of nodes that are adjacent to $z$. For each pair of nodes $v$ and $w$, the number of common neighbors of these two nodes is the number of nodes that are adjacent to both nodes $v$ and $w$.
\begin{equation}
\label{eq:CN}
\displaystyle
\mathcal{CN}(v,w)= |\Lambda(v) \cap \Lambda(w)|
\end{equation}

\item \textit{Resource Allocation} ($\mathcal{RA}$): Zhou et al.~\cite{zhou2009} proposed this measure for addressing link prediction and showed that it provided slightly better performance than $\mathcal{CN}$. This measure assumes that each node has given some resources that will be distributed equally among its neighbors. Then, this idea is adapted by incorporating two nodes $v$ and $w$.
\begin{equation}
\label{eq:RA}
\displaystyle
\mathcal{RA}(v,w) = \sum_{z \in \{\Lambda(v) \cap \Lambda(w)\} \atop z\neq v \neq w}{\frac{1}{|\Lambda(z)|}}
\end{equation}

\item \textit{Adamic-Adar Coefficient} ($\mathcal{AAC}$): Ever since this measure was proposed by Adamic et al.~\cite{adamic2003}, the Adamic-Adar Coefficient has been used in different areas of social network analysis, such as link prediction. The idea behind this measure is to count the common neighbors weighted by the inverse of the logarithm.
\begin{equation}
\label{eq:AAC}
\displaystyle
\mathcal{AAC}(v,w) = \sum_{z \in \{\Lambda(v) \cap \Lambda(w)\} \atop z\neq v \neq w}{\frac{1}{log|\Lambda(z)|}}
\end{equation}

\item \textit{Jaccard Index} ($\mathcal{JI}$): This measure was first proposed in information retrieval ~\cite{Salton1986} as a method for quantifying the similarity between the contents of two sets. This idea is applied to the neighbors of any two nodes as follows:
\begin{equation}
\label{eq:JI}
\displaystyle
\mathcal{JI}(v,w) = \frac{|\Lambda(v) \cap \Lambda(w)|}{|\Lambda(v) \cup \Lambda(w)|} 
\end{equation}

\item \textit{Preferential Attachment} ($\mathcal{PA}$): Newman~\cite{newman2001} showed that in collaboration networks the probability of collaboration between any two nodes (authors) $v$ and $w$ is correlated to the product of $\Lambda(v)$ and $\Lambda(w)$. 
\begin{equation}
\label{eq:PA}
\displaystyle
\mathcal{PA}(v,w) = |\Lambda(v)|\cdot|\Lambda(w)|
\end{equation}

\item \textit{S{\o}rensen-Dice Index} ($\mathcal{SD}$): This measure has been used in ecology to find the similarity between species in ecological data~\cite{dice1945} and it is defined as:
\begin{equation} 
\label{eq:SD}
\displaystyle
\mathcal{SD}(v,w) = \frac{ 2 \times |\Lambda(v) \cap \Lambda(w)|}{|\Lambda(v)| + |\Lambda(w)|} 
\end{equation}

\item \textit{Hub Promoted Index} ($\mathcal{HPI}$): This measure was used to find the similarity between two nodes in a networks with herirarical structures~\cite{ravasz2002}, and it is defined as:
\begin{equation} 
\label{eq:HPI}
\displaystyle
\mathcal{HPI}(v,w) = \frac{ |\Lambda(v) \cap \Lambda(w)|}{min (|\Lambda(v)| , |\Lambda(w)|)} 
\end{equation}

\item \textit{Hub Depressed Index} ($\mathcal{HDI}$): Similar to $\mathcal{HPI}$, the $\mathcal{HDI}$ is defined as:
\begin{equation} 
\label{eq:HDI}
\displaystyle
\mathcal{HDI}(v,w) = \frac{ |\Lambda(v) \cap \Lambda(w)|}{max (|\Lambda(v)| , |\Lambda(w)|)} 
\end{equation}

\item \textit{Local community degree measures} ($\mathcal{CAR}$)\footnote{We stick to the name $\mathcal{CAR}$ as provided by the authors in~\cite{cannistraci2013}.}: Measuring the similarity between two nodes can also be done by looking at how the common neighbors of these two nodes are connected to these two nodes. The common neighbor measure based on the local community degree measure was introduced in ~\cite{cannistraci2013} and is defined as:
\begin{equation} 
\label{eq:CAR}
\displaystyle
\mathcal{CAR}(v,w) = \sum_{z \in \{\Lambda(v) \cap \Lambda(w)\} \atop z\neq v \neq w}{\frac{| \Lambda(v)\cap \Lambda(w) \cap \Lambda(z)|}{|\Lambda(z)|}}
\end{equation}

\end{itemize}
The similarity measures described above are used for undirected networks. For directed networks, two versions of each measure are used by providing two versions of the neighborhood set $\Lambda$, the in-neighbors $\Lambda(v)_{in}$ and the out-neighbors $\Lambda(v)_{out}$. Based on this, an {\it in} and an {\it out} version of the above measures can be constructed. For example, the $\mathcal{CN}_{in}$ for two nodes $v,w$ is: $\mathcal{CN}(v,w)_{in} = |\Lambda(v)_{in} \cap \Lambda(w)_{in}|$.
\subsubsection{Network global features: }
\label{subsec:networkglobalfeatures}
$\ $
Assume that FDM$_{G_i}$ is the feature data model constructed from the network $G_i$, then we call FDM${_\mathcal{G}}$ the aggregated model from all networks $\mathcal{G}$. For FDM${_\mathcal{G}}$, network global features are required to represent the global properties of each network $G_i$. This means that a pair of nodes $v,w$ appears $|\mathcal{G}|$ times in the combined FDM${_\mathcal{G}}$. These global features help the classification algorithm discriminate among different instances of $v,w$ if their edge proximity features are close to each other. Therefore, it is crucial to label the instances in the FDM${_\mathcal{G}}$ with network global features. Network density is used as a network global feature.
\textit{Network density} ($\eta$): Is a measure that reflects the degree of completeness of a network, and it is defined as:
\begin{equation}
\label{eq:density}
\displaystyle
\eta({G_i}) = \frac{2\cdot|E(G_i)|}{|V(G_i)|\cdot(|V(G_i)|-1)} 
\end{equation}

\begin{figure}[h]
\label{fig:proposedsolutionfigure}
  \centering
    \includegraphics[scale=1]{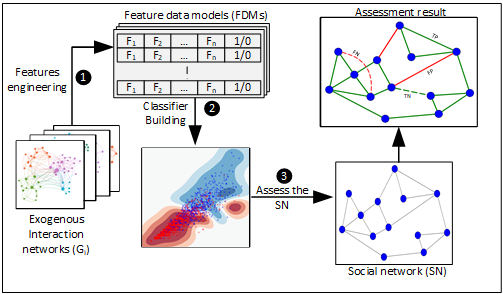}
  \caption {(Color online) The framework for link assessment using associated interaction networks and machine learning. 
}
  \label{fig:proposedMethod}
\end{figure}

Based on the above description, the FDM${_\mathcal{G}}$ for an undirected network contains a number of instances that is equal to $\sum_{G_i \in \mathcal{G}}{\frac{|V(G_i)|\cdot(|V(G_i)|-1)}{2}}$ such that for every pair of nodes $v,w \in V(G_i)$ and $G_i \in \mathcal{G}$, an instance $\mathcal{I}(v,w)$ is a tuple that contains: (1) the edge proximity features' values for $v$ and $w$ presented in Equations~\ref{eq:CN} to \ref{eq:CAR}; (2) the network global feature of $\mathcal{G}$ presented in Equation~\ref{eq:density}; (3) a binary class, \{${1,0}$\}, which indicates whether there is a link $e=\{v,w\}$ in $G_i$ or not. This binary class is what we are predicting here.\\
Figure~\ref{fig:proposedMethod} depicts the process of assessing the links of a social network $SN$ using associated interaction networks $\mathcal{G}$. In step 1, the FDM$_{G_i}$ is constructed for each network $G_i \in \mathcal{G}$. In step 2, the constructed FDMs are used to train a machine learning classifier which is used, in step 3, to assess the $SN$ by providing the binary classification value. In step 4, the ground truth labels are used to evaluate the classification performance. Based on this method, training and testing are done on two disjoint sets, except when training and testing on the $SN$, which enhances and supports the results, as we will see later in Section~\ref{sec:Results}.

\section{Ground truth and experiment setup}
\label{sec:groundtruth}
Let the ground truth $SN=(V,E)$ be the network with the set of nodes $V$ and with the edge set $E$ that contains only true-positives and true-negatives.
Let $SN_{predicted}=(V,E')$ be the predicted social network on the same set of nodes $V$ and the set of edges $E'$ is the predicted edges. Accordingly, the set $E- E'$ contains the false-negative links, i.e., links that exist in reality (in $E$) but were not observed in the $SN_{predicted}$ (in $E'$). Similarly, $E'-E$ contains the false-positive links, i.e., those that do not exist in $SN$ but were observed in the $SN_{predicted}$. The goal is now to get a classification result that is as close to $SN$ as possible. Therefore, the more accurate the machine learning classifier, the more efficient the link assessment method.\\
Based on that, the machine learning problem \textit{$\psi(X,Y)$} means that the data $X$ is used to train a machine learning classifier to classify the links in $Y$, where $X \neq Y$.
To test the effectiveness of this method, a social network with ground truth data will be assessed.
If the links of the social network $SN$ are assessed using one network $G_i \in \mathcal{G}$, then the machine learning problem becomes: \textit{$\psi$(FDM$_{G_i},SN$)}, which means that the training phase uses the FDM generated only from a single network $G_i$ to assess the links in (to test on) the $SN$. In this case, global network feature is excluded, as it is fixed for all instances of the same network, and thus is useless for the classifier algorithm. This assessment enables us to determine whether the structure of a network $G_i \in \mathcal{G}$ is sufficient to efficiently assess the links in the $SN$ or not. Additionally, this will provide insights regarding the correlation between this single network and the social network. Similarly, if the links of the $SN$ are assessed using the whole set of the interaction networks, then the machine learning problem becomes: \textit{$\psi$(FDM$_{\mathcal{G}},SN$)}, which means that the training phase uses the aggregated FDMs of all interaction networks, and in this case the global network feature is included.

In order to test the proposed solution, the following experiment steps were performed:
\begin{enumerate}
\item \textit{Build the FDM}: In this step the values of the features described in Section~\ref{fdm} were calculated, which constructs the $FDM_{G_i}$ for every network $G_i$ of the interaction networks $\mathcal{G}$, and also for the social network of interest $FDM_{SN}$, where $SN$ is the ground truth to test on.
\item \textit{Training and testing}: The classifier was trained using different training sets depending on the goal of the experiment. To assess the links of the social network of interest $SN$ using a single network $G_i \in \mathcal{G}$, the training set was $FDM_{G_i}$ and the test set was the $FDM_{SN}$. For assessing the links of the social network of interest $SN$ using the whole set of the associated interaction networks, the training set was $FDM_{\mathcal{G}}$ and the test set was the $FDM_{SN}$. To assess the links of the social network of interest $SN$ using the $SN$ itself, training and testing were done using the $FDM_{SN}$ with $k$-fold cross-validation.
\item \textit{Evaluation}: The evaluation metrics described in Section~\ref{sec:evaluationmetrics} were used to evaluate the classification results of the training sets.
\end{enumerate}

\section{Datasets and evaluation metrics}
\label{sec:datasets}
In this section, a description of the datasets and the evaluation metrics will be presented.
\subsection{Datasets}
In order to validate the proposed method, we tested it using two different social networks with their associated interaction networks.
The first dataset ($RG$) was the research group dataset described in Section~\ref{introduction}. The social network for the research group~\cite{raey2013}, which is the Facebook social network, was considered as the ground truth online social network for its members\footnote{This may sound contradictory to what we claimed in the introduction concerning noise in social networks. However, we contacted the owner of the dataset and made sure that there are neither false-positives nor false-negatives in the Facebook network.}. All the networks of the first dataset are undirected. The second dataset ($LF$) was a \textit{law firm} dataset~\cite{Emmanuel2012} containing an offline directed social network along with the following two exogenous interaction networks based on questionnaires:
\begin{itemize}
\item \textit{Advice} $G_1$: If a member $A$ seeks advice from another member $B$, then there is a directed link from $A$ to $B$.
\item \textit{Cowork} $G_2$: If a member $A$ considers another member $B$ a co-worker, then there is a directed link from $A$ to $B$.
\end{itemize}
The \textit{Friendship} network of the law firm dataset and the \textit{Facebook} social network of the research group dataset are considered as the ground truth social networks. That is because both networks were validated by the collectors and the edges in both networks are true-positives and the edges that are absent are true-negatives\footnote{A description of the law firm dataset and how it was collected can be found \href{https://www.stats.ox.ac.uk/~snijders/siena/Lazega_lawyers_data.htm}{on the original publisher page}.}.
Table~\ref{tab:datasetdescription} shows the network statistics of the networks of the datasets used in this paper. These statistics include the number of nodes $n$, the number of links $m$, the average clustering coefficient $cc(G)$, and the network's density $\eta$.

\begin{small}
\begin{table}[htbp]
  \centering
  \caption{Datasets statistics.}
    \begin{tabular}{|c|l|c|c|c|c|}
    \hline
    Dataset & Networks & $n$     & $m$     &  $cc(G_i)$    & $\eta(G_i)$ \\
    \hline
    \multirow{3}[10]{*}{$RG$} & \textbf{$SN$: Facebook}   & $32$    & 248   & 0.48  & 0.5 \\
\cline{2-6}          & {$G_1$: Work}     & 60    & 338   & 0.34  & 0.19 \\
\cline{2-6}          & {$G_2$: Co-author}    & 25    & 42    & 0.43  & 0.14 \\
\cline{2-6}          & {$G_3$: Lunch}    & 60    & 386   & 0.57  & 0.21 \\
\cline{2-6}          & {$G_4$: Leisure}    & 47    & 176   & 0.34  & 0.16 \\
    \hline\hline
    \multirow{2}[6]{*}{$LF$} & \textbf{$SN$: Friends}   & 69    & 339   & 0.43  & 0.07 \\
\cline{2-6}          & $G_1$: Co-work    & 71    & 726   & 0.41  & 0.15 \\
\cline{2-6}          & $G_2$: Advice   & 71    & 717   & 0.42  & 0.14 \\
    \hline
    \end{tabular}%
  \label{tab:datasetdescription}%
\end{table}%
\end{small}

\subsection{Evaluation metrics}
\label{sec:evaluationmetrics}
In order to evaluate the prediction results, we present a set of classical classification evaluation metrics used for evaluating the classification results of the experiment. For any two nodes $v$ and $w$, a true-positive (TP) classification instance means that there is a link $e=\{v,w\}$ between these two nodes in the test set, for example the links of the $SN$, and the classifier succeeds in predicting this link. A true-negative (TN) instance means that there is no link $e=\{v,w\}$ in the test set and the classifier predicts that this link does not exist. On the other hand, a false-negative (FN) instance means that for a pair of nodes $v$ and $w$ there is a link $e=\{v,w\}$ in the test set and the classifier predicts that there is no link. Similarly, false-positive (FP) instance means that for a pair of nodes $v$ and $w$ there is no link $e=\{v,w\}$ in the test set but the classifier predicts that there is a link.\\
Based on the basic metrics described above, we used the following additional evaluation metrics:
\begin{itemize}

\item \textbf{Precision ($\mathcal{P}$)}: the number of true-positives in relation to all positive classifications. It is defined as: $\mathcal{P} = \frac{TP}{TP+FP}$
\item \textbf{Recall ($\mathcal{R}$)}: also called True Positive rate or \textit{Sensitivity}. It is defined as: $\mathcal{R} = \frac{TP}{TP+FN}$
\item \textbf{Accuracy ($\mathcal{ACC}$)}: the percentage of correctly classified instances. It is defined as: $\mathcal{ACC} = \frac{TP+TN}{TP+TN+FP+FN}$
\item \textbf{F-measure ($\mathcal{F}$)}: the harmonic mean of precision and recall. It is defined as: $\mathcal{F} = \frac{2 \cdot \mathcal{P} \cdot \mathcal{R}}{\mathcal{P}+\mathcal{R}}$
\end{itemize}

Those measures, particularly the accuracy, are not informative if there are imbalanced datasets where one class, the no-edge class in the $FDM$, comprises the majority of the dataset instances. To achieve more rigorous validation of the results, we used the weighted version of the above measures to reflect on informative and accurate measures.\\
\textbf{Area under Receiver Operating Characteristics curve} (\textit{AU-ROC}): The ROC~\cite{hanley1982} curve plots the true positive rate against the false positive rate. The area under this curve reflects how good a classifier is and is used to compare the performance of multiple classifiers.

\section{Results}
\label{sec:Results}
In this section, the properties of the constructed $FDMs$ and the classification results will be presented.
\subsection{The properties of the FDM}
\label{subsec:thePropertiesOftheFDM}
\begin{figure}[!ht]
    \centering
    \begin{subfigure}[b]{0.3\textwidth}
    	\centering
        \includegraphics[width=\linewidth]{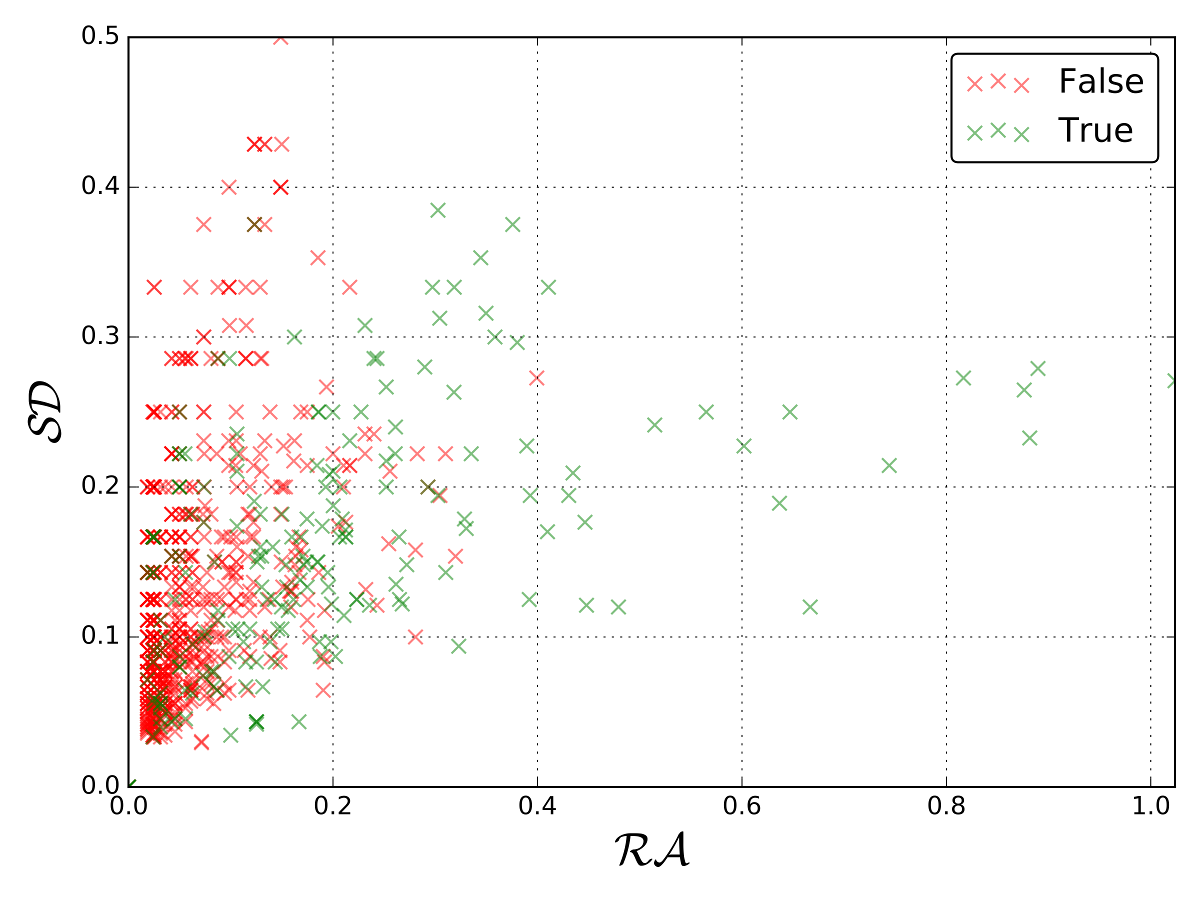}
        \caption{Work}
        \label{fig:FDM_Work}
    \end{subfigure}
    \begin{subfigure}[b]{0.3\textwidth}
        	\centering
        \includegraphics[width=\linewidth]{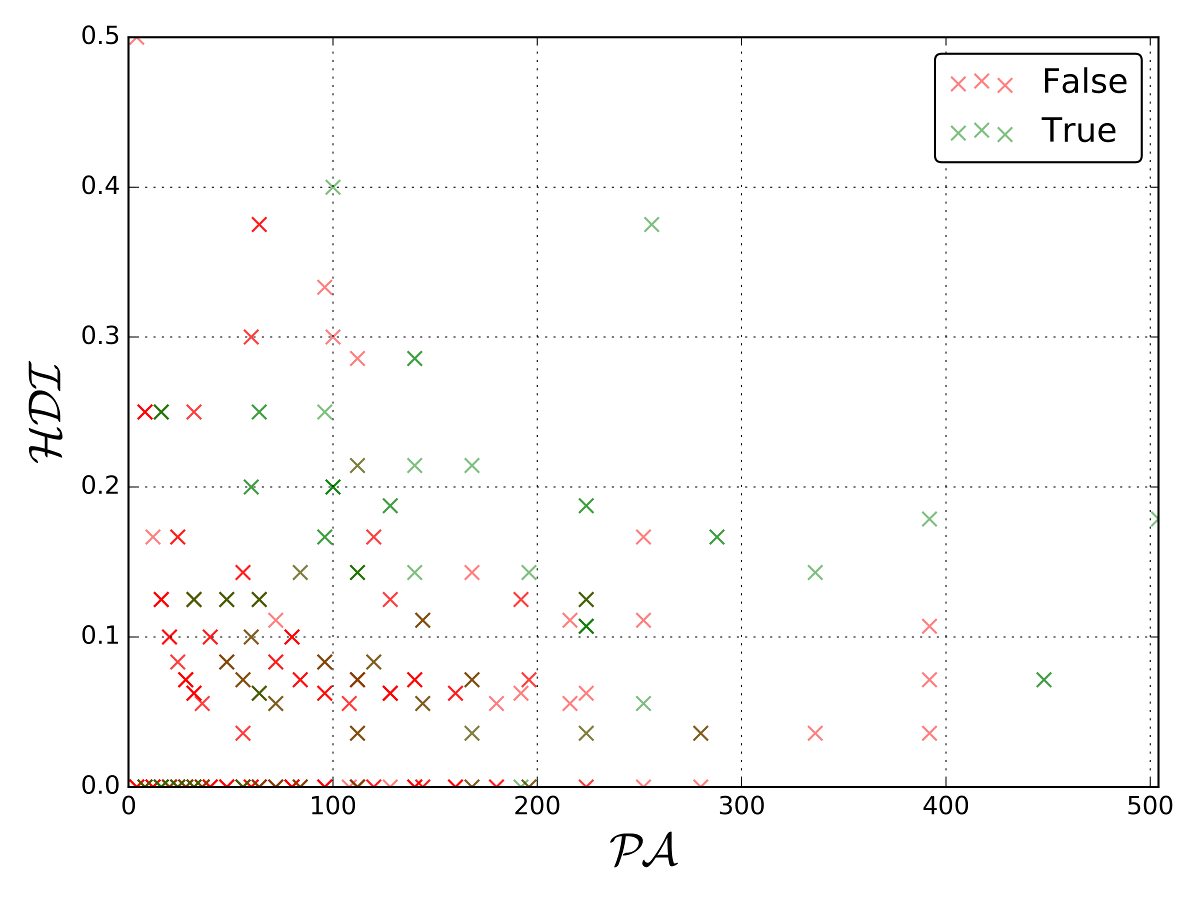}
        \caption{Leisure}
        \label{fig:FDM_Leisure}
    \end{subfigure}
    \begin{subfigure}[b]{0.3\textwidth}    
    	\centering
        \includegraphics[width=\linewidth]{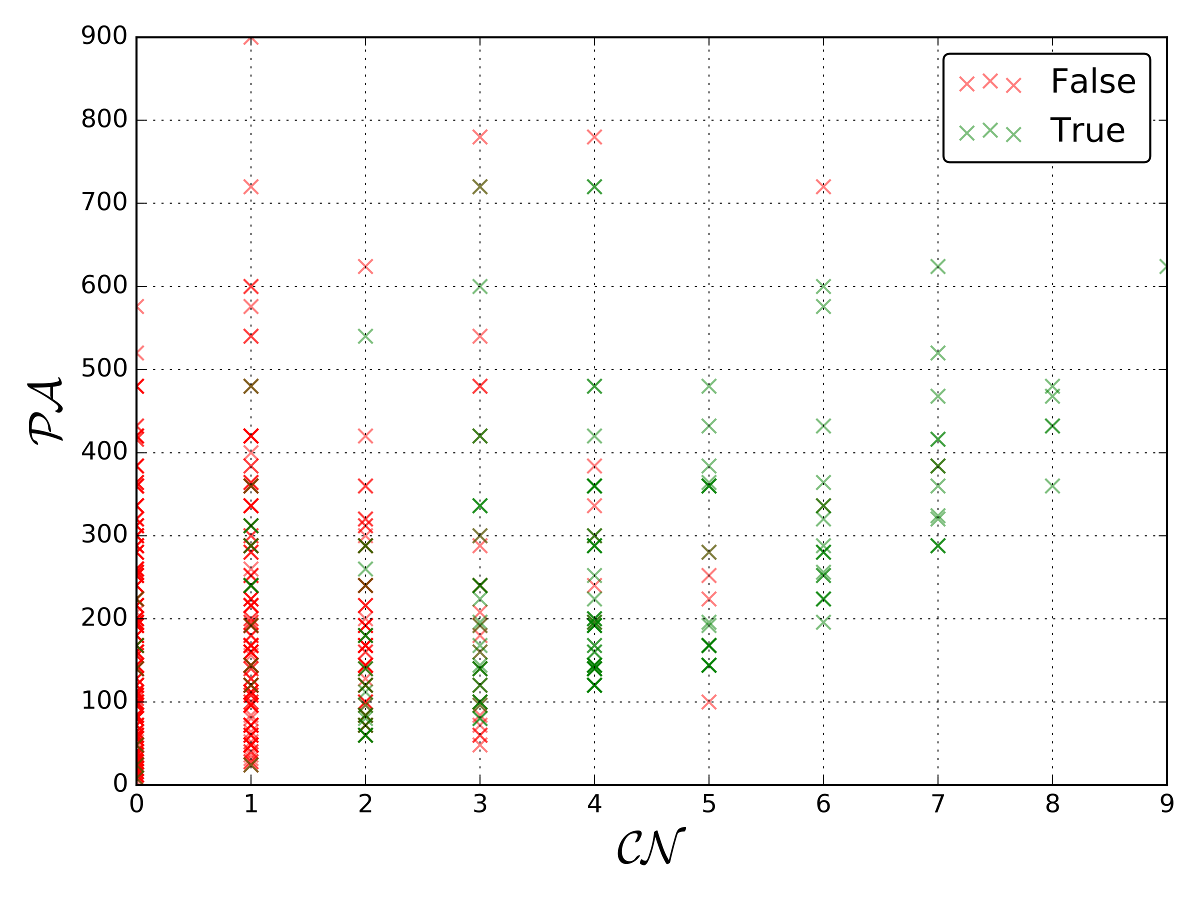}
        \caption{Lunch}
        \label{fig:FDM_Lunch}
    \end{subfigure}
    \begin{subfigure}[b]{0.3\textwidth}
        	\centering
        \includegraphics[width=\linewidth]{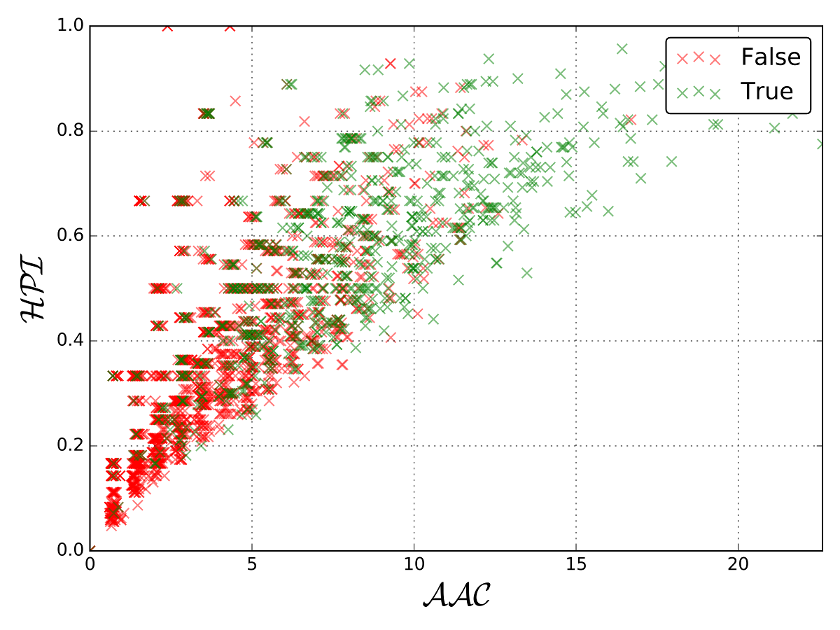}
        \caption{Advice}
        \label{fig:FDM_work}
    \end{subfigure}
    \begin{subfigure}[b]{0.3\textwidth}
        	\centering
        \includegraphics[width=\linewidth]{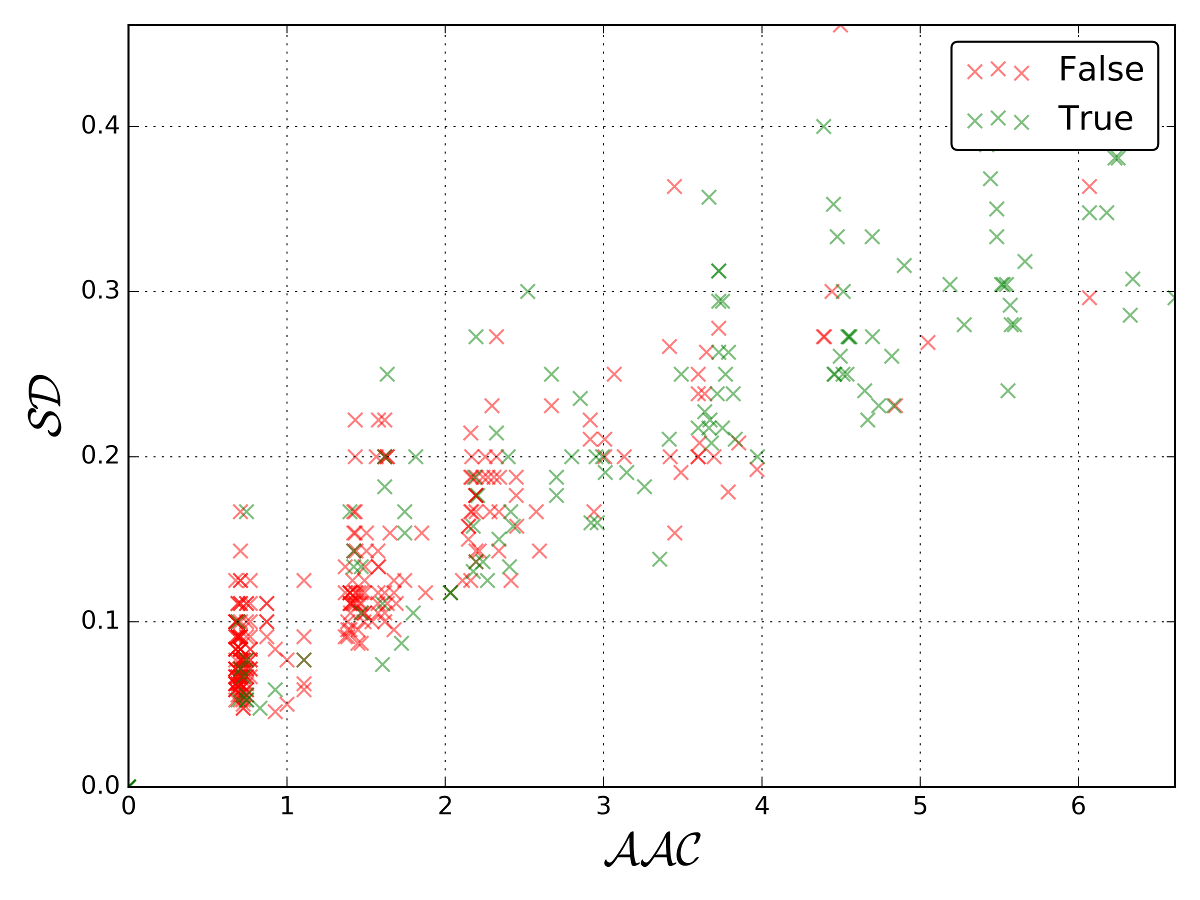}
        \caption{Facebook}
        \label{fig:FDM_aggregated_rg}
    \end{subfigure}
    \begin{subfigure}[b]{0.3\textwidth}
        	\centering
        \includegraphics[width=\linewidth]{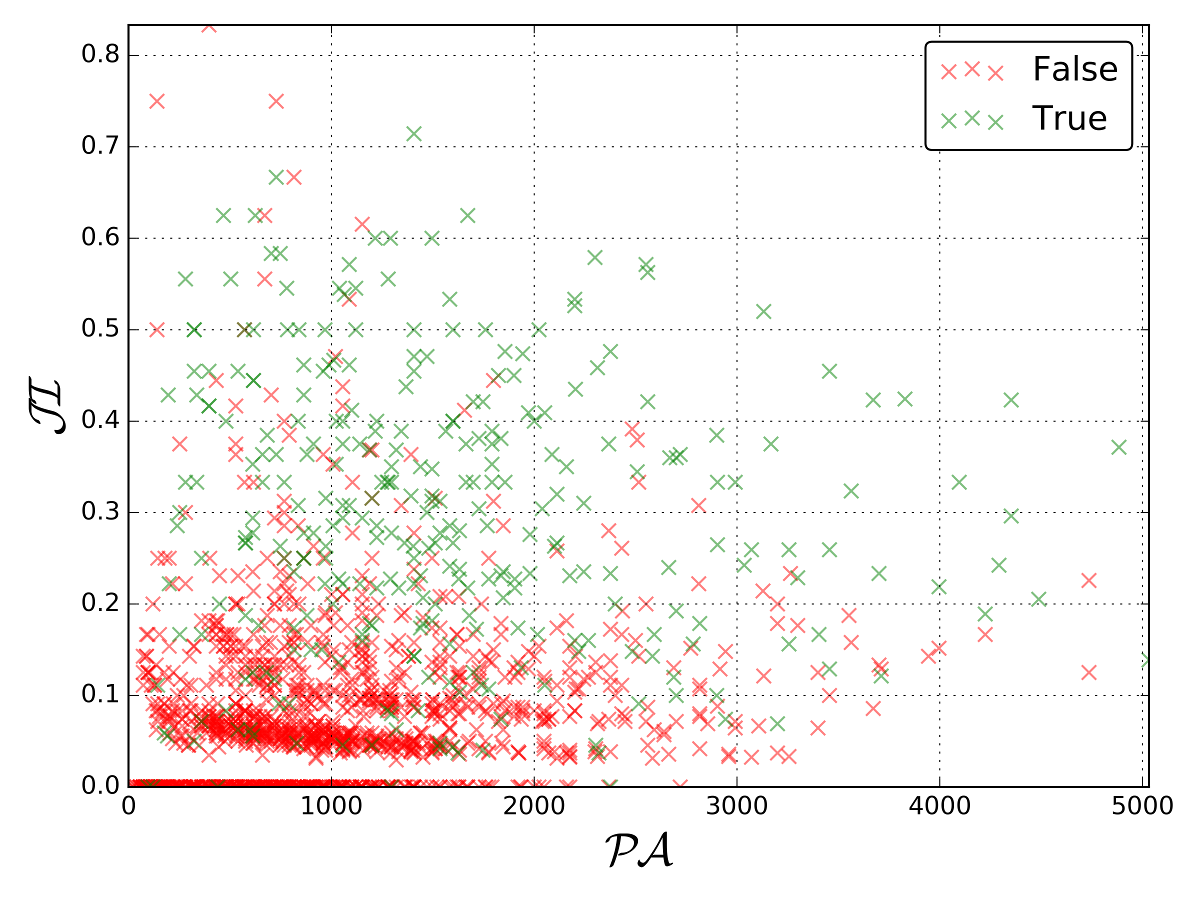}
        \caption{Aggregated RG}
        \label{fig:FDM_advice}
    \end{subfigure}
        \begin{subfigure}[b]{0.3\textwidth}
            	\centering
        \includegraphics[width=\linewidth]{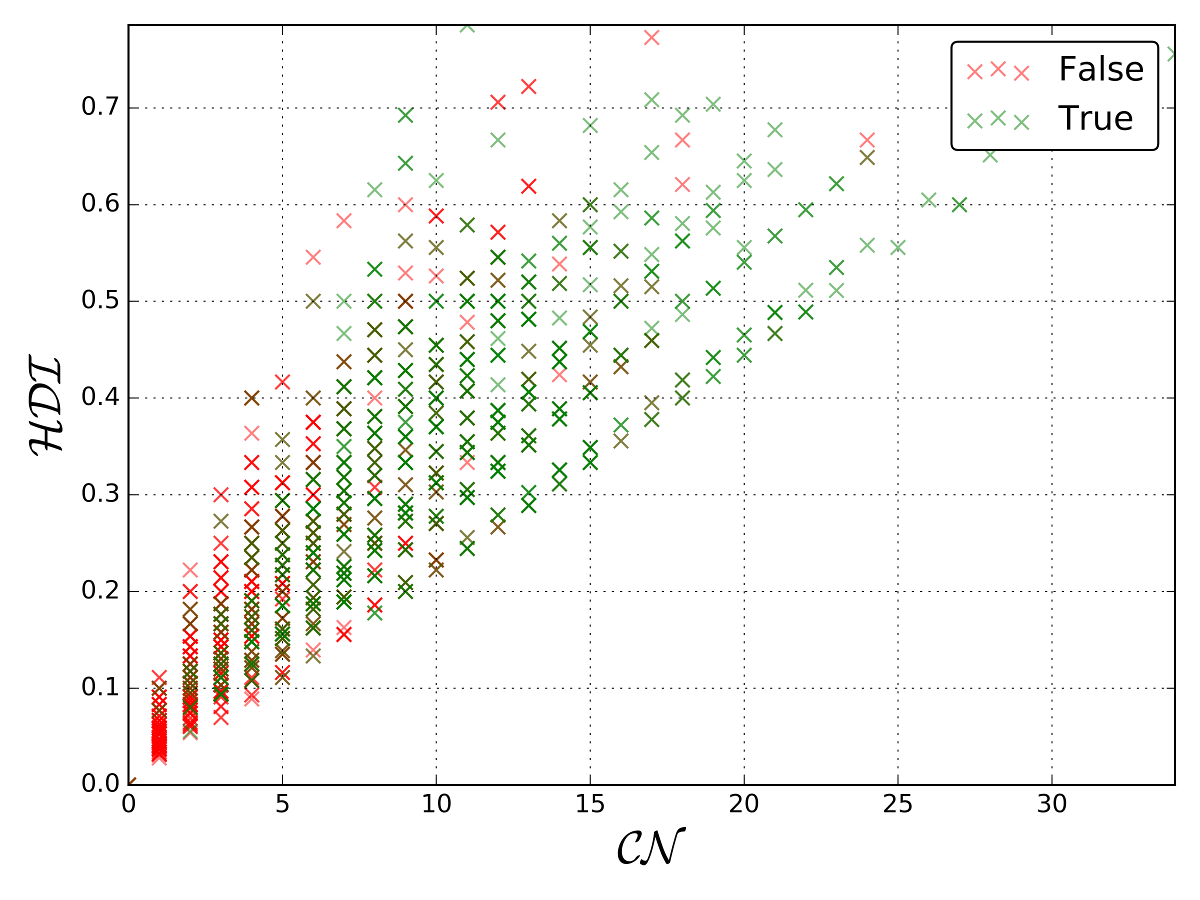}
        \caption{Coworker}
        \label{fig:FDM_coworker}
    \end{subfigure}
      \begin{subfigure}[b]{0.3\textwidth}
            	\centering
        \includegraphics[width=\linewidth]{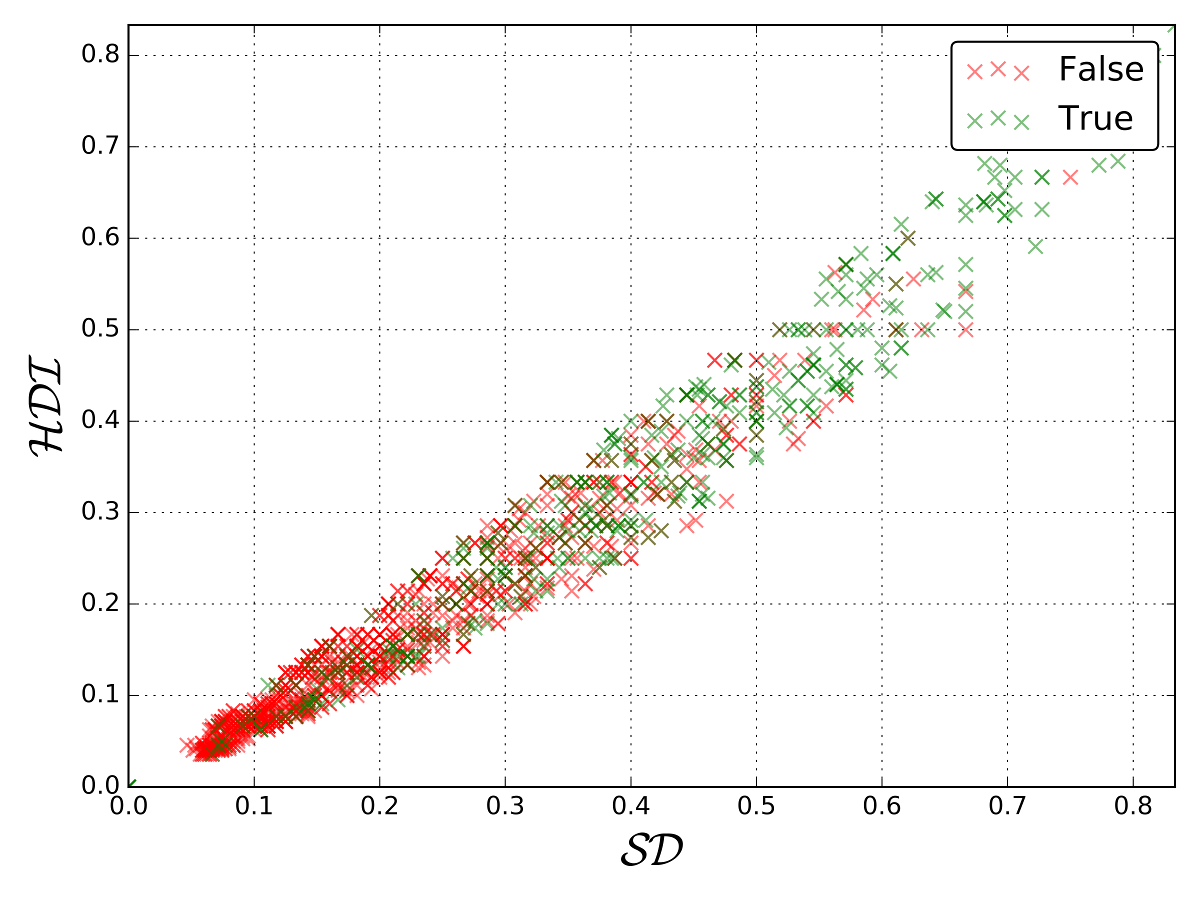}
        \caption{Friendship}
        \label{fig:FDM_friend}
    \end{subfigure}
        \begin{subfigure}[b]{0.3\textwidth}
            	\centering
        \includegraphics[width=\linewidth]{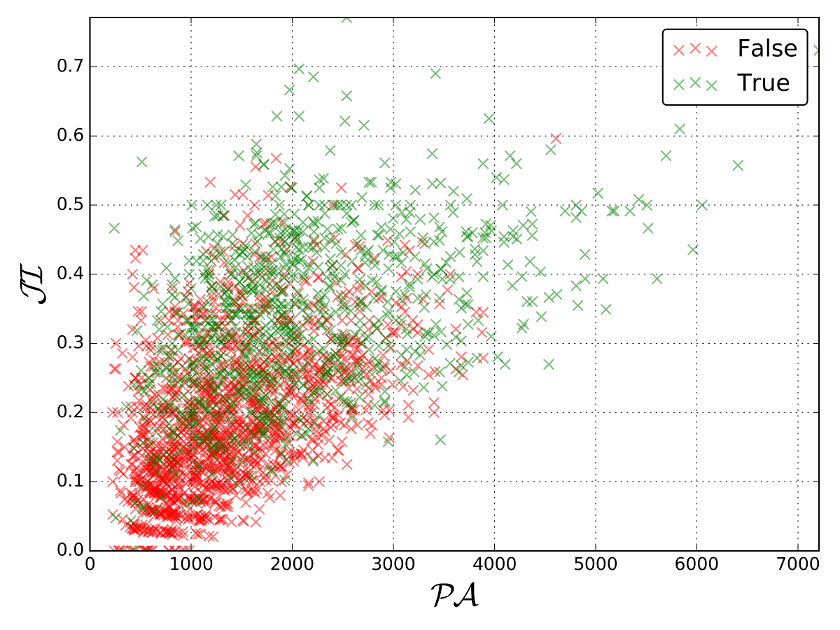}
        \caption{Aggregated LF}
        \label{fig:FDM_aggregated_lf}
    \end{subfigure}
    \caption{(Color online) Selected 2-D scatters of the $FDM$ for the used networks. The x-axis and the y-axis represent selected features presented in Equations~\ref{eq:CN} to~\ref{eq:CAR}. The red markers are the \textit{False} instances and green markers are the \textit{True} instances which indicate the existence and the non-existence of an edge, respectively. }
     \label{fig:nonLinearlySeparable}
\end{figure}

In this section, some properties of the $FDM$'s feature and what they look like will be presented.
Figure~\ref{fig:nonLinearlySeparable} shows a selected two dimensions(2-D) of the FDMs constructed from the used networks.
The figure shows that the $FDM$ is not linearly separable, which renders the classification problem non-trivial for linear classification models. The figure shows also that there are some features that are highly correlated, for example, Figure~\ref{fig:FDM_friend} shows a strong correlation between the $\mathcal{SD}$ and the $\mathcal{HDI}$ features. Later, we discuss the correlation between the features and their impact on the classification process.

\begin{figure}[!ht]
    \centering
    \begin{subfigure}[b]{0.32\textwidth}
    	\centering
        \includegraphics[width=\linewidth]{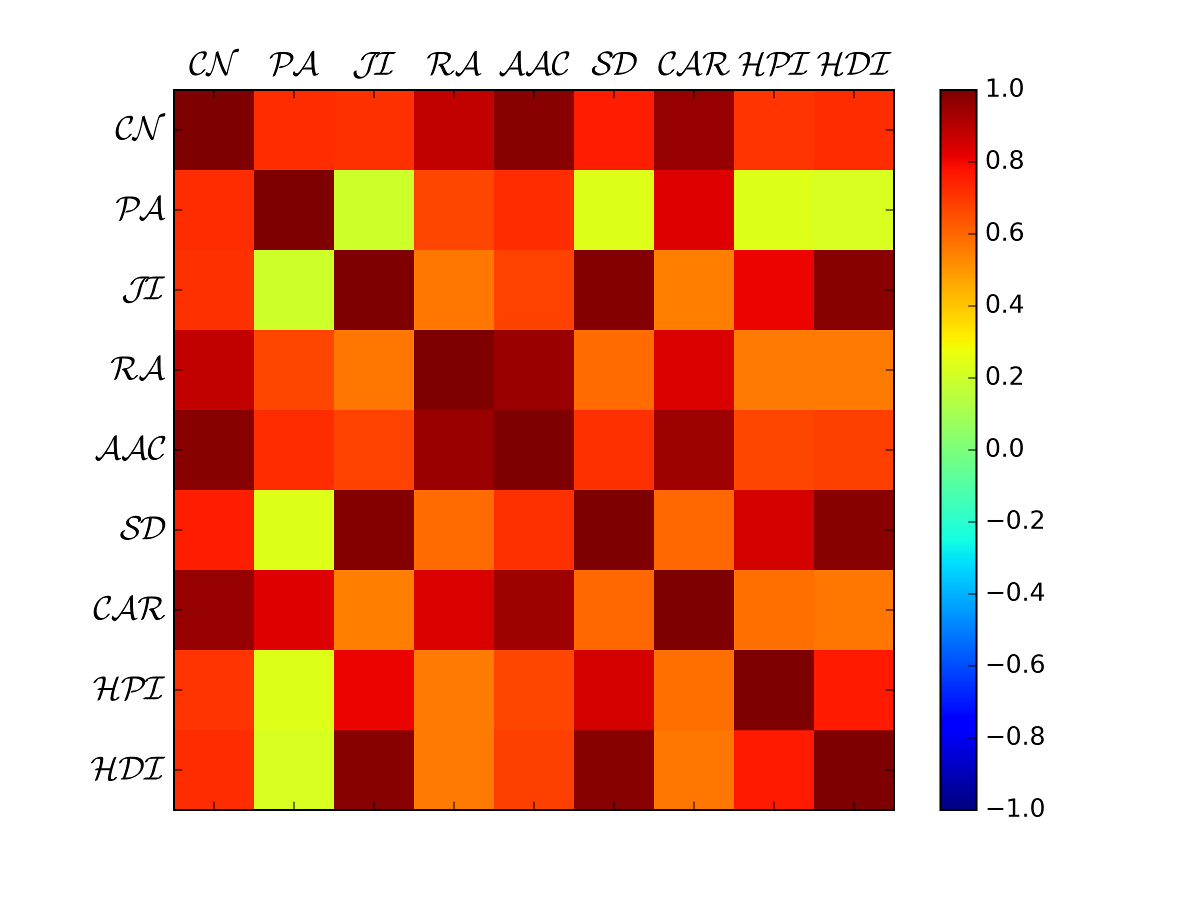}
        \caption{}
        \label{fig:RG_FDM_a}
    \end{subfigure}\hspace{0em}
    \begin{subfigure}[b]{0.32\textwidth}
        	\centering
        \includegraphics[width=\linewidth]{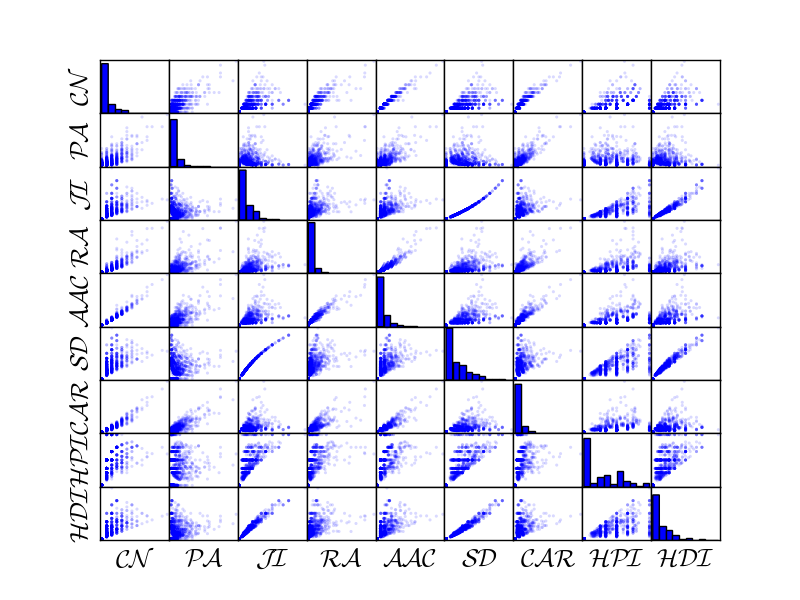}
        \caption{}
        \label{fig:RG_FDM_b}
    \end{subfigure}\hspace{0em}
    \begin{subfigure}[b]{0.32\textwidth}    
    	\centering
        \includegraphics[width=\linewidth]{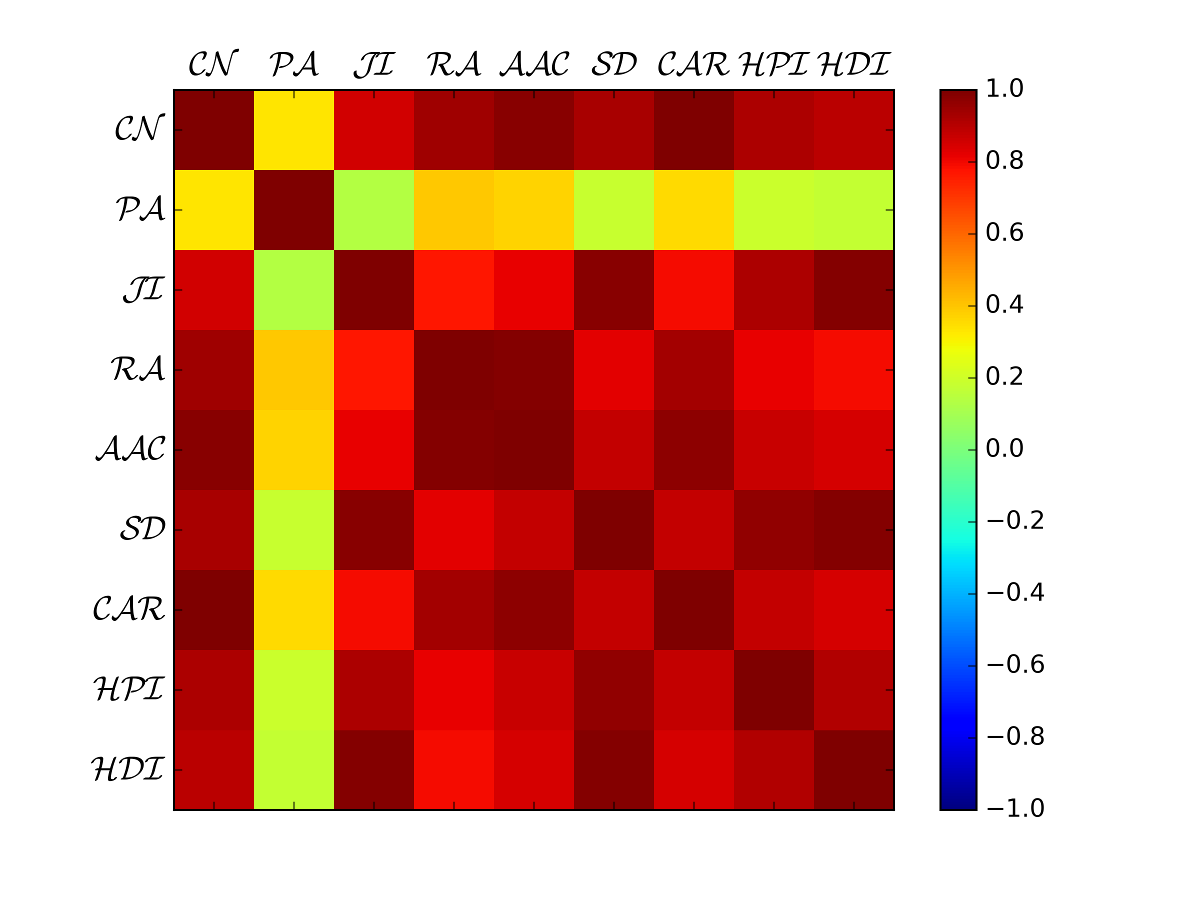}
        \caption{}
        \label{fig:RG_FDM_c}
    \end{subfigure}\hspace{0em}
    \begin{subfigure}[b]{0.32\textwidth}
        	\centering
        \includegraphics[width=\linewidth]{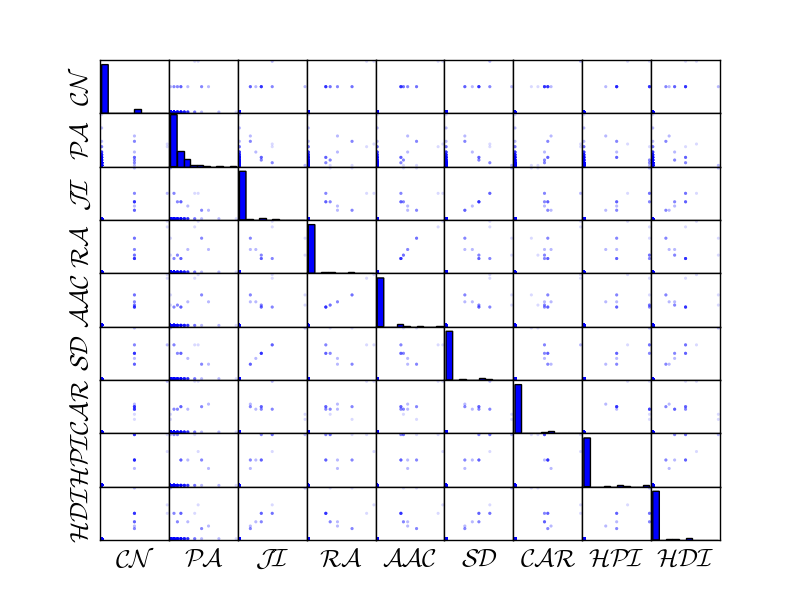}
        \caption{}
        \label{fig:RG_FDM_d}
    \end{subfigure}\hspace{0em}
    \begin{subfigure}[b]{0.32\textwidth}    
    	\centering
        \includegraphics[width=\linewidth]{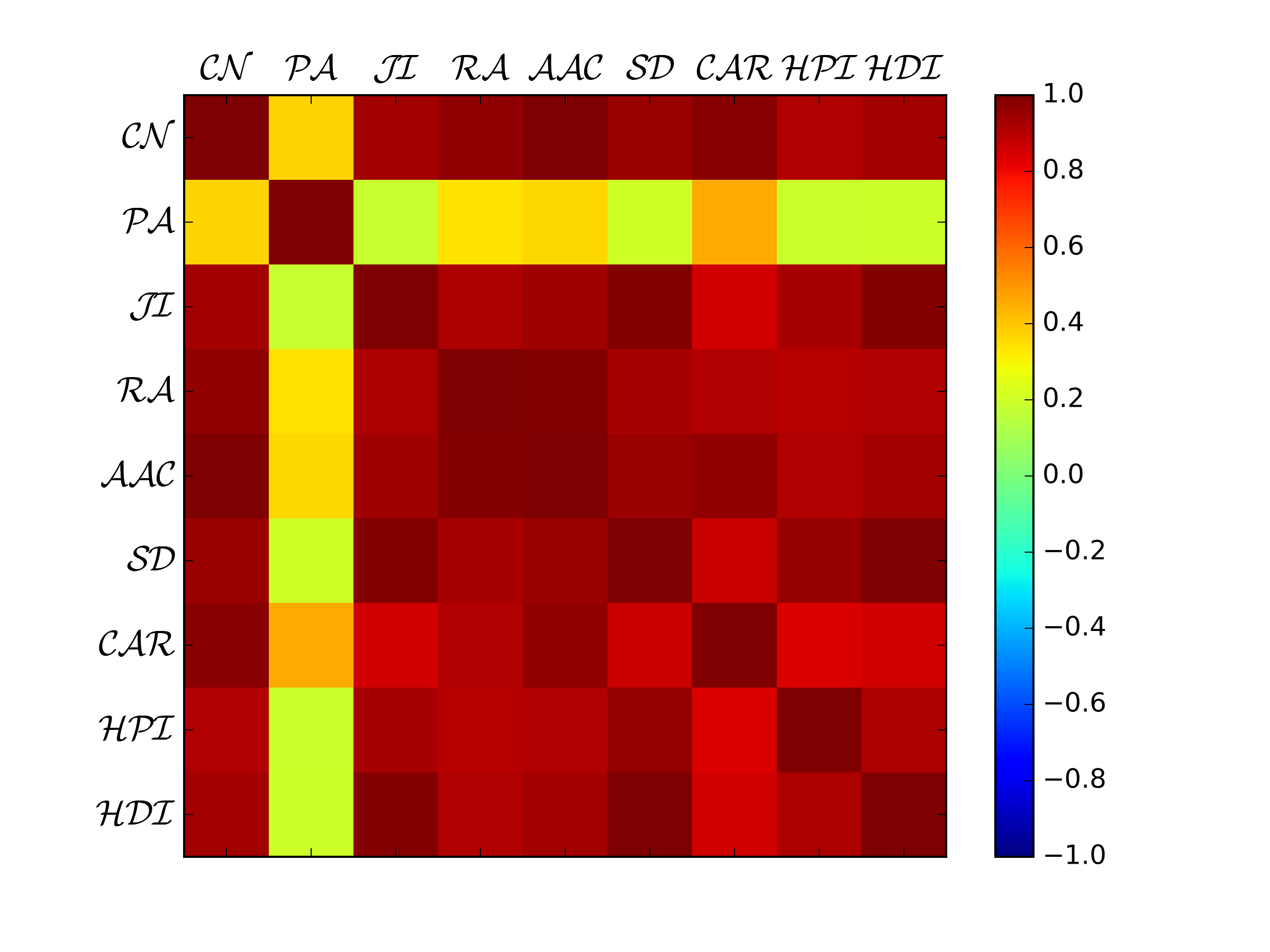}
        \caption{}
        \label{fig:RG_FDM_e}
    \end{subfigure}\hspace{0em}
    \begin{subfigure}[b]{0.32\textwidth}
        	\centering
        \includegraphics[width=\linewidth]{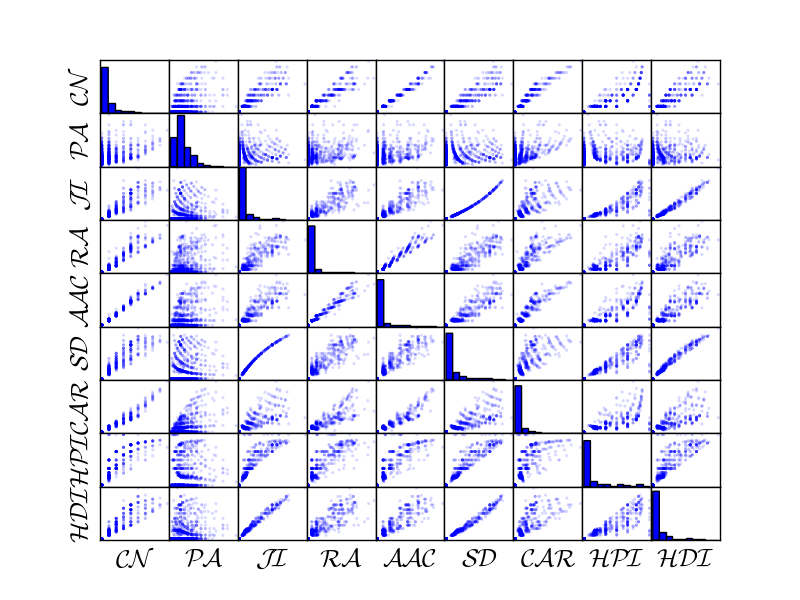}
        \caption{}
        \label{fig:RG_FDM_f}
    \end{subfigure}\hspace{0em}
     \begin{subfigure}[b]{0.32\textwidth}    
    	\centering
        \includegraphics[width=\linewidth]{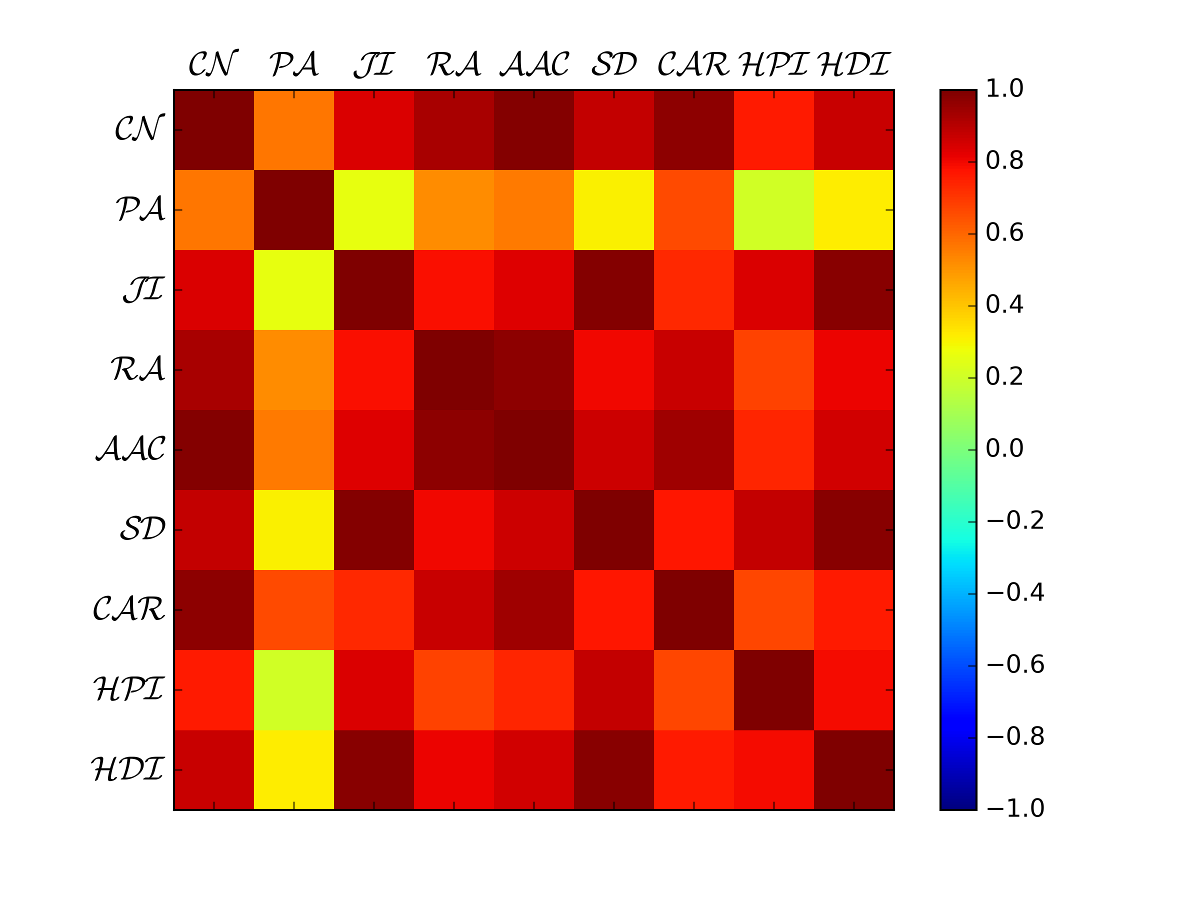}
        \caption{}
        \label{fig:RG_FDM_g}
    \end{subfigure}\hspace{0em}
    \begin{subfigure}[b]{0.32\textwidth}
        	\centering
        \includegraphics[width=\linewidth]{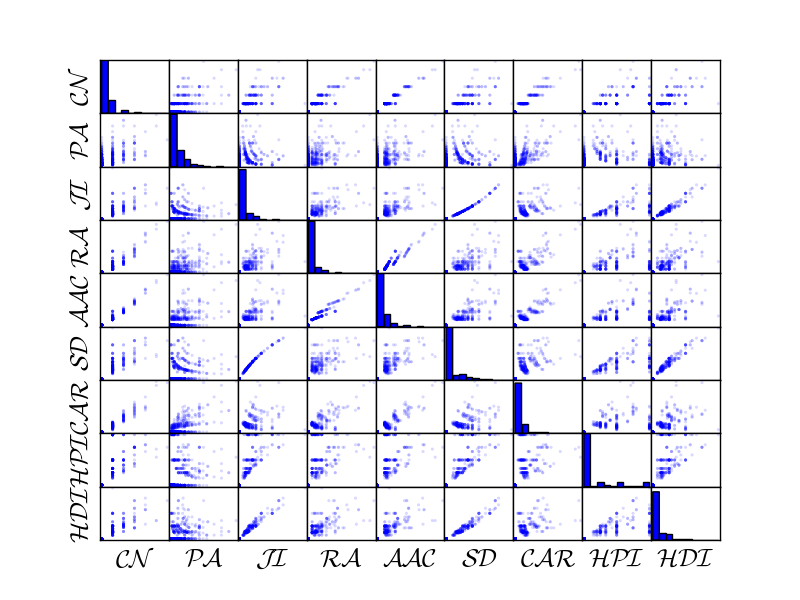}
        \caption{}
        \label{fig:RG_FDM_h}
    \end{subfigure}\hspace{0em}
     \begin{subfigure}[b]{0.32\textwidth}    
    	\centering
        \includegraphics[width=\linewidth]{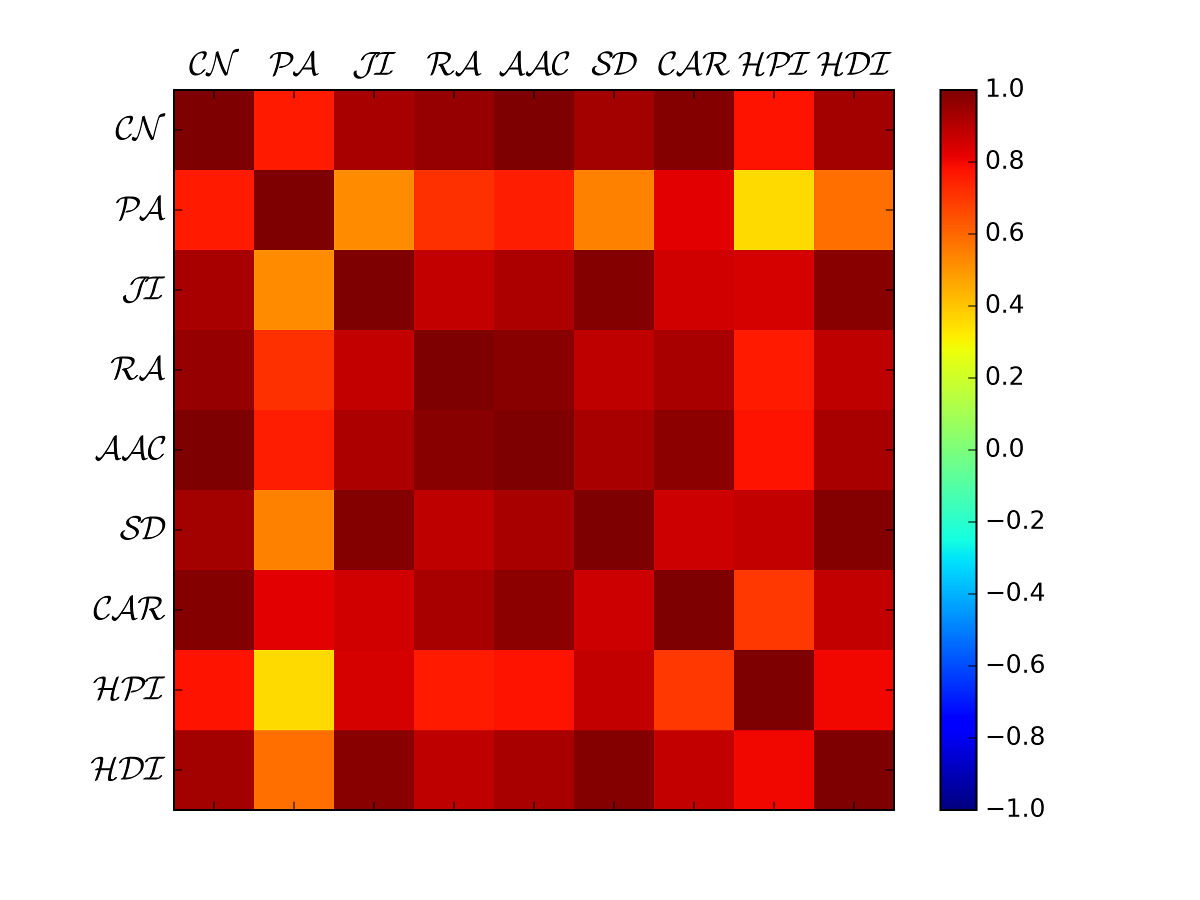}
        \caption{}
        \label{fig:RG_FDM_i}
    \end{subfigure}\hspace{0em}
    \begin{subfigure}[b]{0.32\textwidth}
        	\centering
        \includegraphics[width=\linewidth]{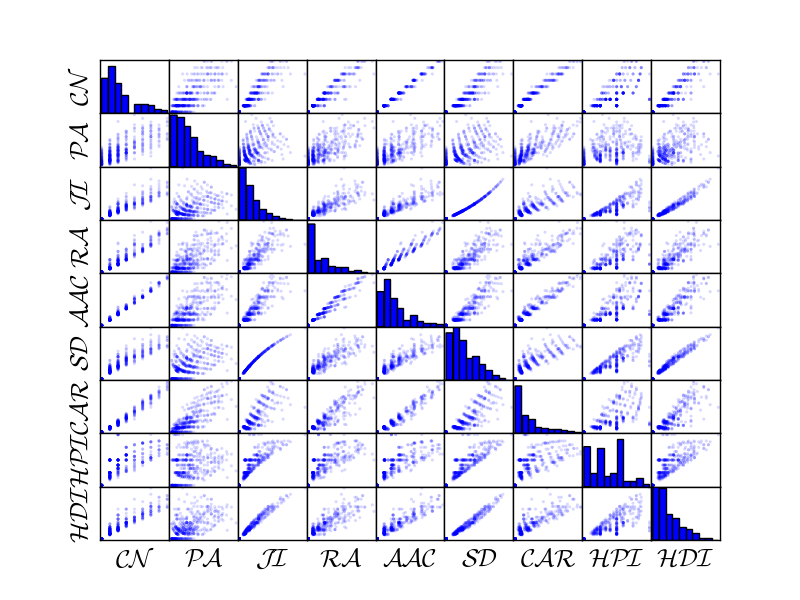}
        \caption{}
        \label{fig:RG_FDM_j}
    \end{subfigure}\hspace{0em}
         \begin{subfigure}[b]{0.32\textwidth}    
    	\centering
        \includegraphics[width=\linewidth]{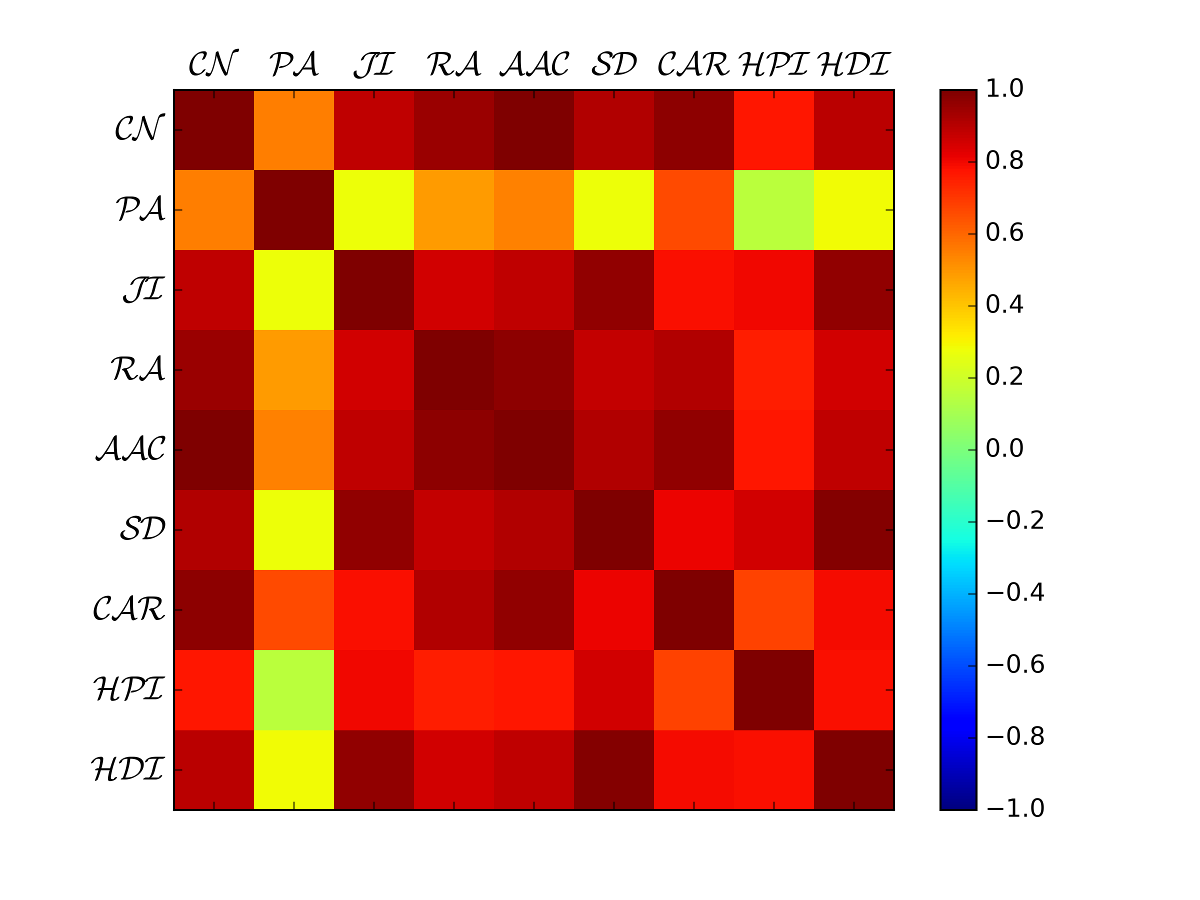}
        \caption{}
        \label{fig:RG_FDM_k}
    \end{subfigure}\hspace{0em}
    \begin{subfigure}[b]{0.32\textwidth}
        	\centering
        \includegraphics[width=\linewidth]{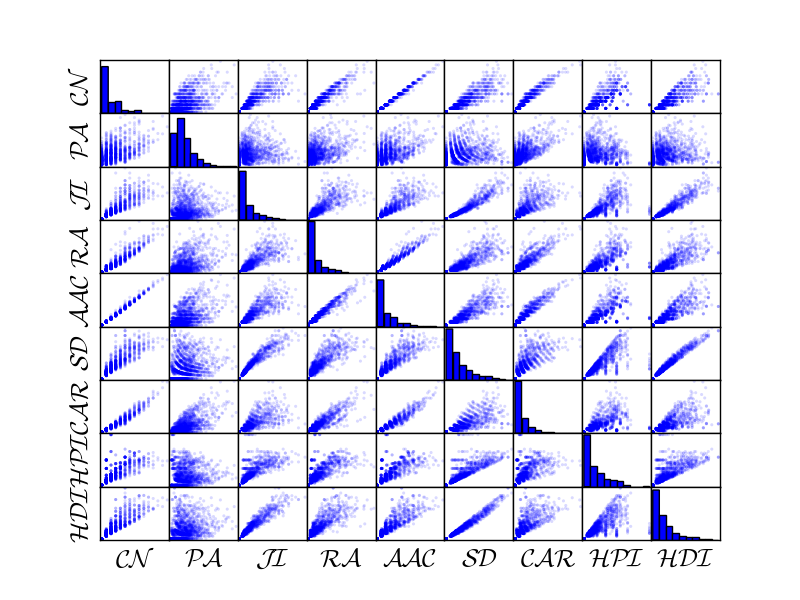}
        \caption{}
        \label{fig:RG_FDM_l}
    \end{subfigure}
     \caption{(Color online) The feature correlation matrix and the feature correlation scatter of the features of the $FDM$ for the Research Group (RG) dataset. Panels a,c,e,g,i, and k show the correlation matrix for the $FDM$ of the networks Work, Coauthor, Lunch, Leisure, Facebook, and Aggregated RG, respectively. Panels b,d,f,h,j, and l show the correlation scatter between two feature of the $FDM$ for those networks, with the distribution of each feature in the diagonal. }
     \label{fig:RG_FDM}
\end{figure}

\begin{figure}[!ht]
    \centering
    \begin{subfigure}[b]{0.32\textwidth}
    	\centering
        \includegraphics[width=\linewidth]{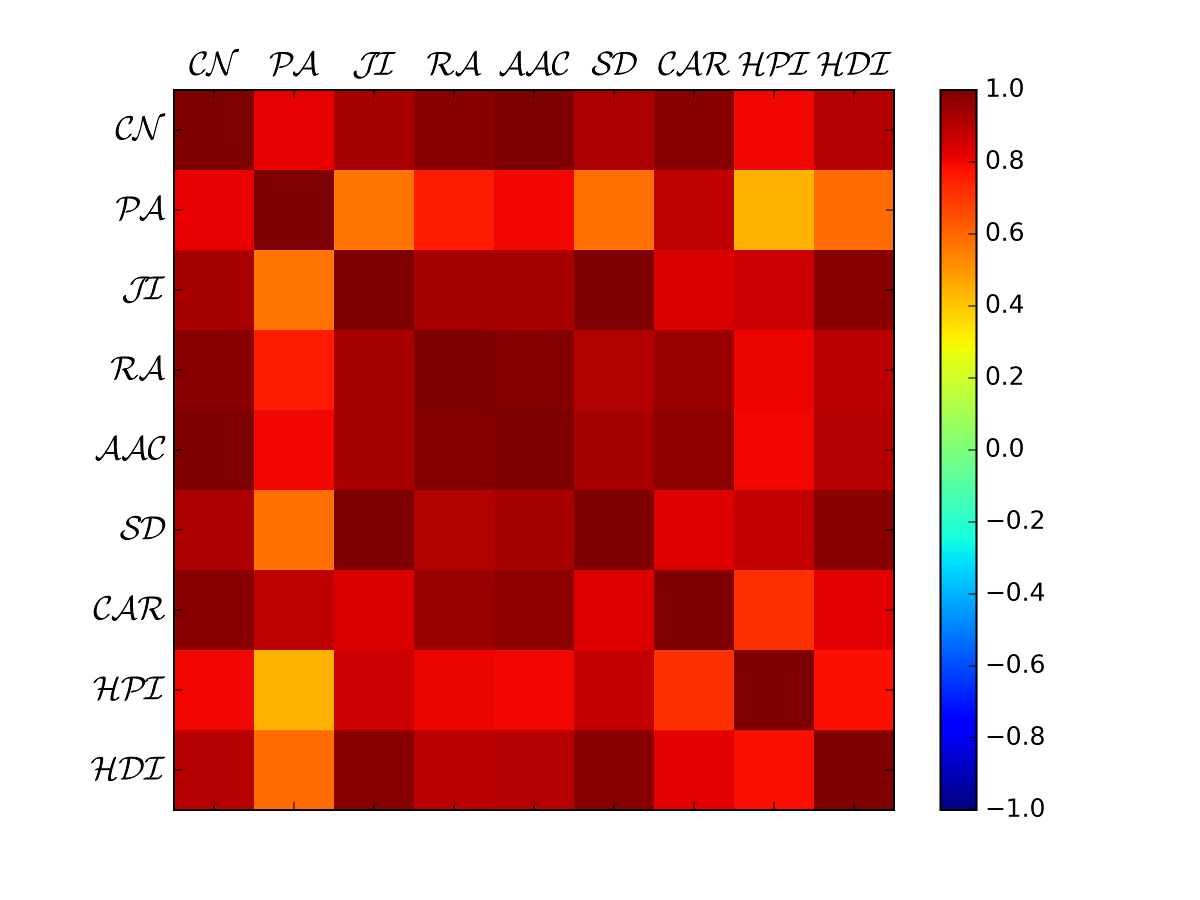}
        \caption{}
        \label{fig:matrix_advice}
    \end{subfigure}\hspace{0em}
    \begin{subfigure}[b]{0.32\textwidth}
        	\centering
        \includegraphics[width=\linewidth]{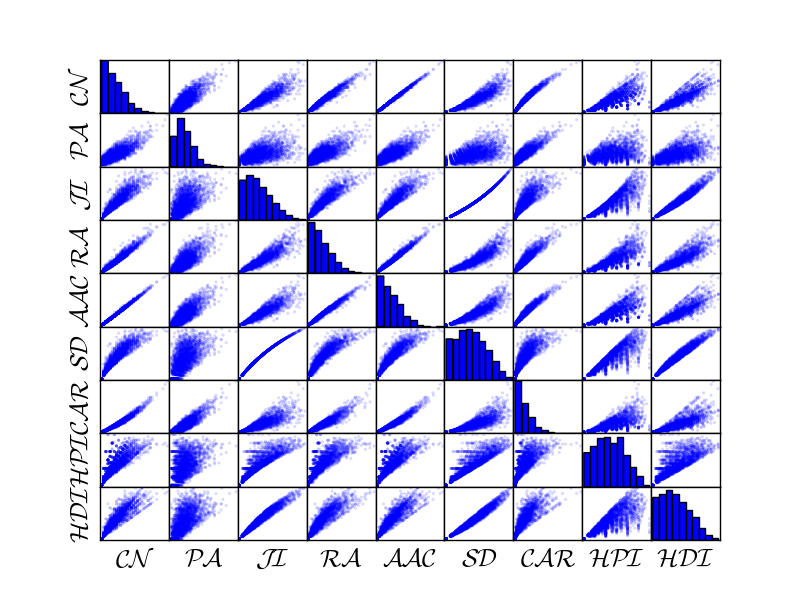}
        \caption{}
        \label{fig:scatter_advice}
    \end{subfigure}\hspace{0em}
    \begin{subfigure}[b]{0.32\textwidth}    
    	\centering
        \includegraphics[width=\linewidth]{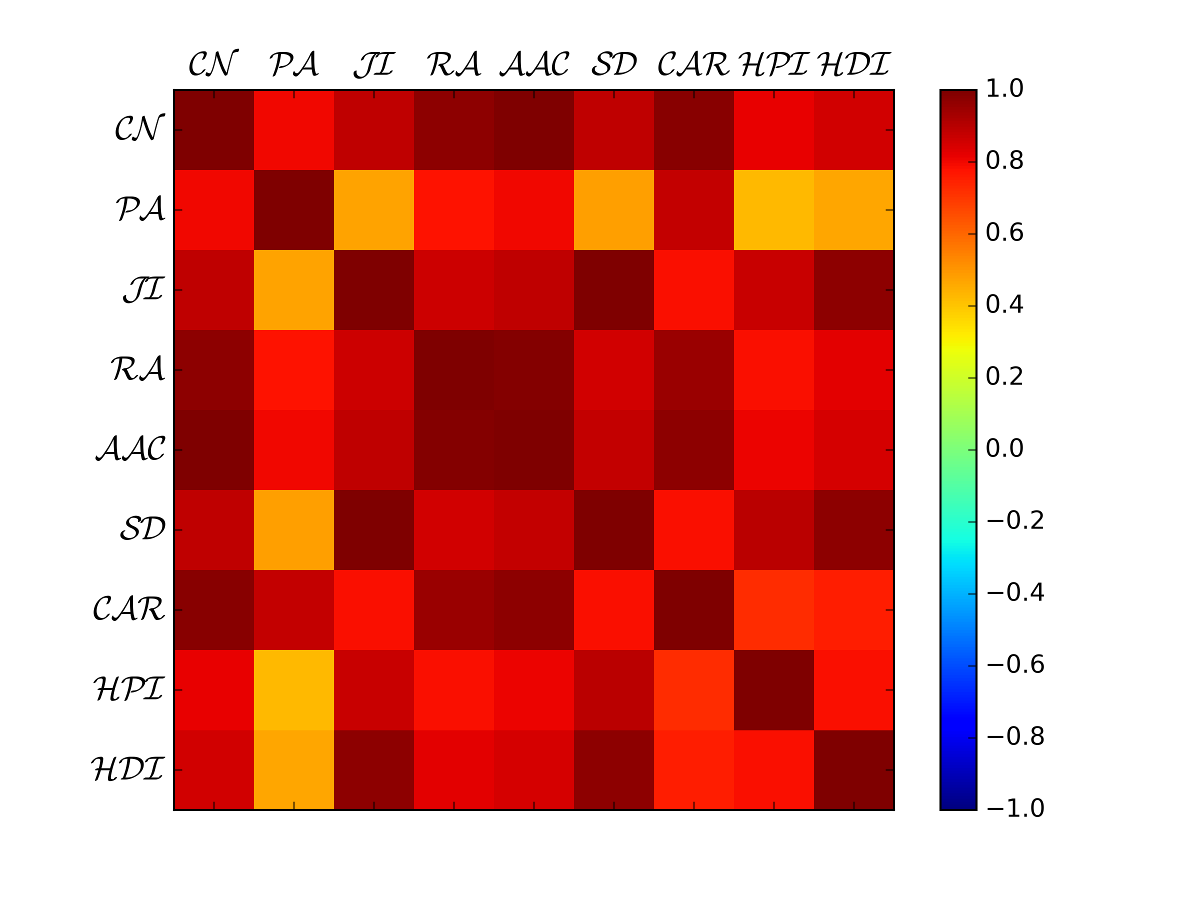}
        \caption{}
        \label{fig:matrix_coworker}
    \end{subfigure}\hspace{0em}
    \begin{subfigure}[b]{0.32\textwidth}
        	\centering
        \includegraphics[width=\linewidth]{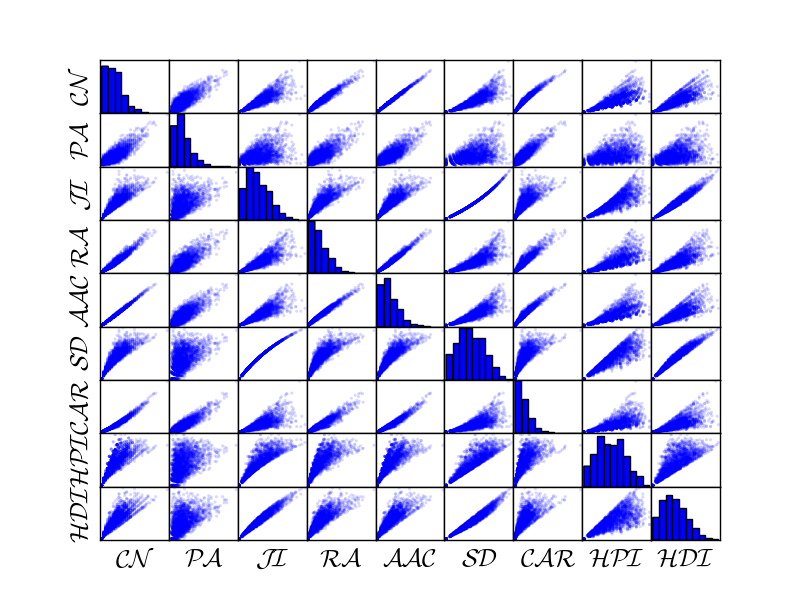}
        \caption{}
        \label{fig:scatter_coworker}
    \end{subfigure}\hspace{0em}
    \begin{subfigure}[b]{0.32\textwidth}    
    	\centering
        \includegraphics[width=\linewidth]{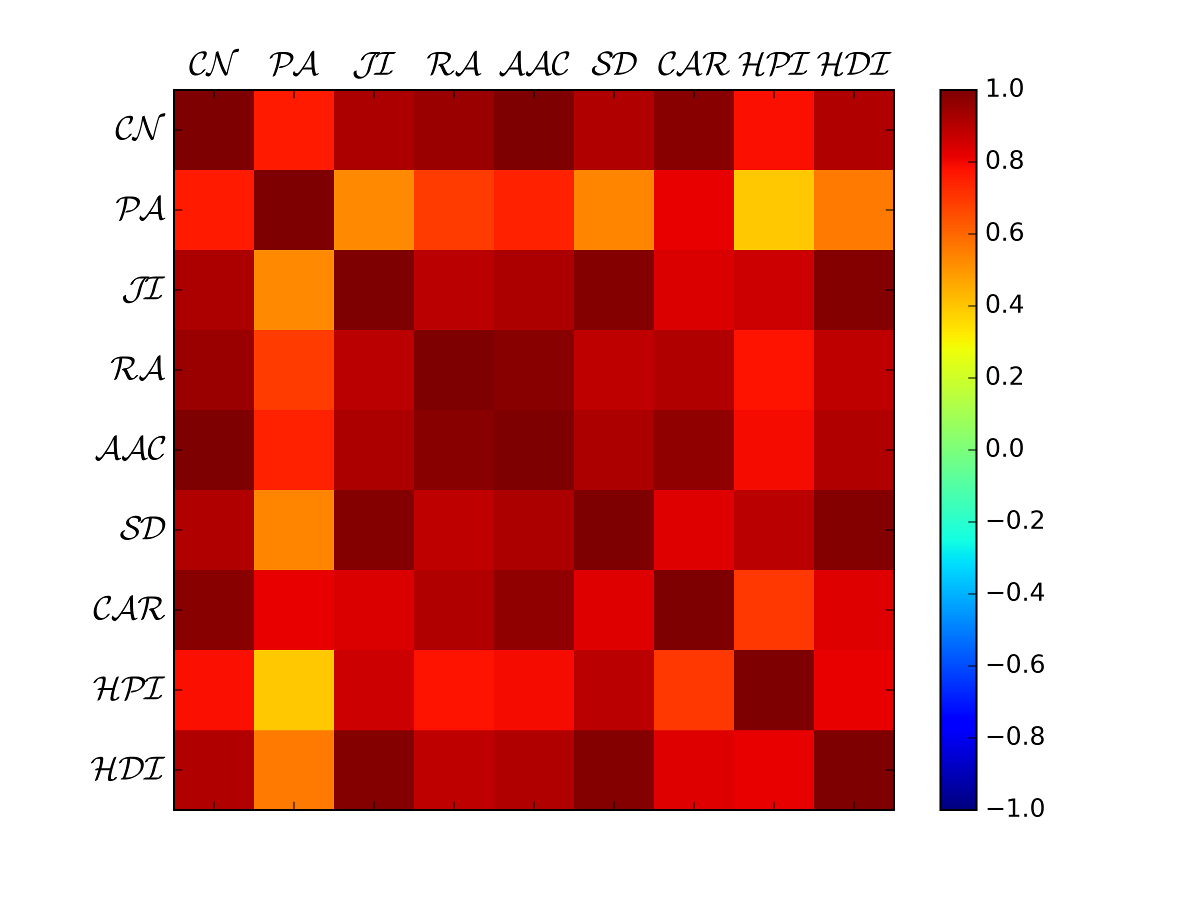}
        \caption{}
        \label{fig:matrix_friend}
    \end{subfigure}\hspace{0em}
    \begin{subfigure}[b]{0.32\textwidth}
        	\centering
        \includegraphics[width=\linewidth]{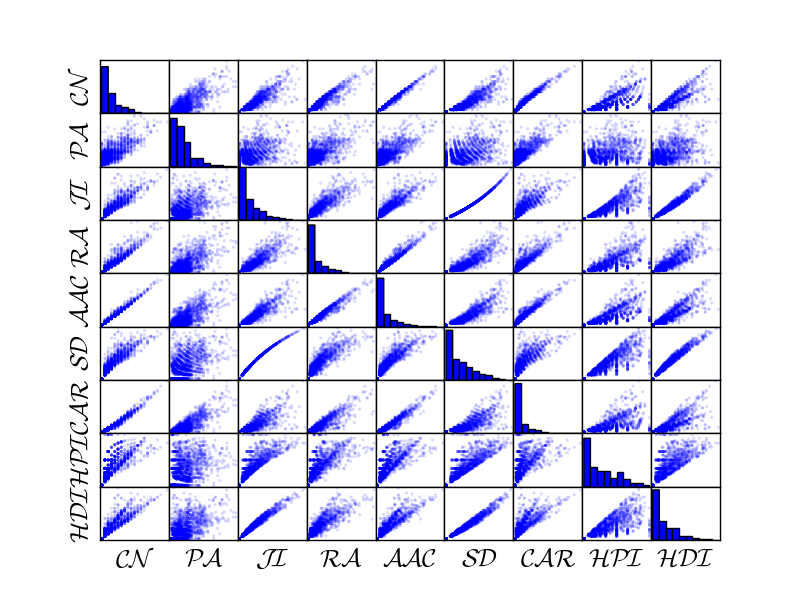}
        \caption{}
        \label{fig:scatter_friend}
    \end{subfigure}\hspace{0em}
         \begin{subfigure}[b]{0.32\textwidth}    
    	\centering
        \includegraphics[width=\linewidth]{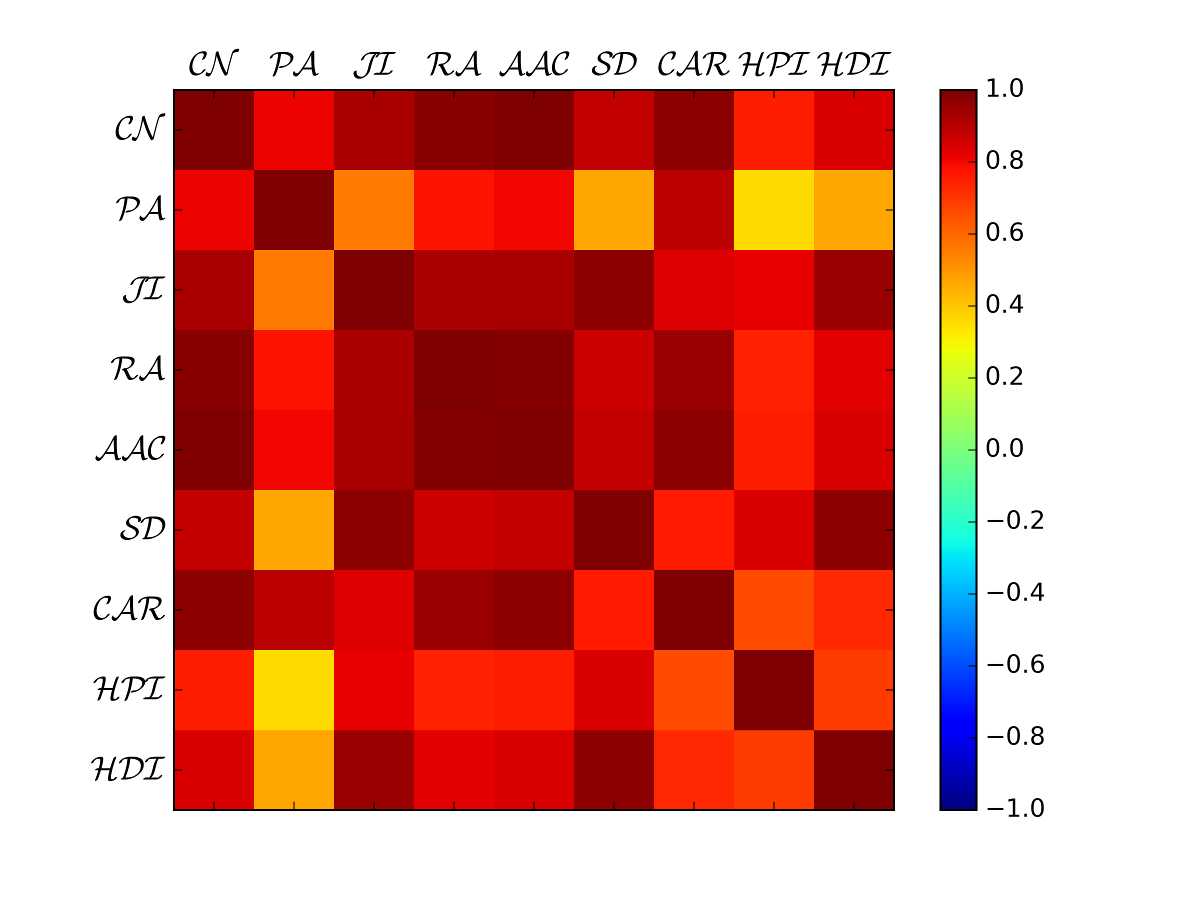}
        \caption{}
        \label{fig:matrix_aggregated_lf}
    \end{subfigure}\hspace{0em}
    \begin{subfigure}[b]{0.32\textwidth}
        	\centering
        \includegraphics[width=\linewidth]{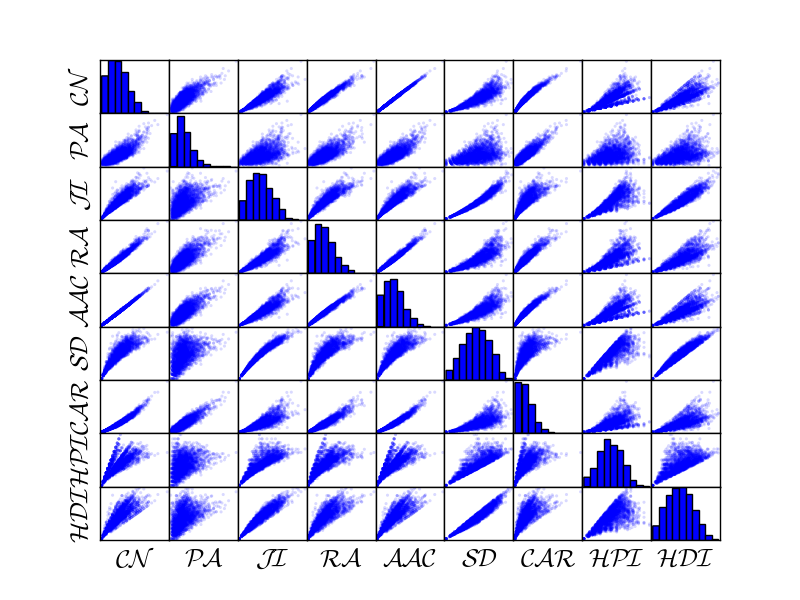}
        \caption{}
        \label{fig:scatter_aggregated_lf}
    \end{subfigure}
     \caption{(Color online) The feature correlation matrix and the feature correlation scatter of the $FDM$'s feature for the Law Firm (LF) dataset. Panels a,c,e, and g show the correlation matrix for the $FDM$ of the networks Advice, Coworker, Friend, and Aggregated LF, respectively. Panels b,d,f,and h show the correlation scatter between two features each of the $FDM$ for those networks, with the distribution of each feature in the diagonal.}
     \label{fig:LF_FDM}
\end{figure}

There are many machine learning classifiers, each with its own assumptions, limitations, and parameters to tune. For example, some classifiers like \textit{Logistic Regression} assumes that there is no correlation between the features. This makes logistic regression not suited for classification with correlated features. Whereas, there are classifiers, such as \textit{Support Vector Classifier} with kernels, which can perform well with correlated features; others assume a Gaussian distribution of the features, and so on. Thus, it is crucial to understand the data that is being used in the classification process. Figures~\ref{fig:RG_FDM} and~\ref{fig:LF_FDM} shows deeper analysis of the $FDM$'s features. In Figure~\ref{fig:RG_FDM}, the correlation between the features of the $FDM$ is not the same across all networks of the \textit{Research Group} dataset. In Figure~\ref{fig:RG_FDM_a} (the Work network), there is less correlation between the features when compared with, for example, Figure~\ref{fig:RG_FDM_i} (Facebook). From Figure~\ref{fig:RG_FDM}, panels~\ref{fig:RG_FDM_a},~\ref{fig:RG_FDM_c},~\ref{fig:RG_FDM_e},~\ref{fig:RG_FDM_g},~\ref{fig:RG_FDM_i}, and~\ref{fig:RG_FDM_k}, the feature that is correlated the least with the other features is the $\mathcal{PA}$. It turned out that the FDM's features are intrinsically correlated. The reason is that, unlike the other features, the $\mathcal{PA}$ feature is not dependent in the $\mathcal{CN}$. The correlation is clearer in the corresponding correlation scatters in Figure~\ref{fig:RG_FDM}, panels~\ref{fig:RG_FDM_b},~\ref{fig:RG_FDM_d},~\ref{fig:RG_FDM_f},~\ref{fig:RG_FDM_h},~\ref{fig:RG_FDM_j}, and~\ref{fig:RG_FDM_l}. These panels show a strong correlation between $\mathcal{JI}$ and $\mathcal{SD}$, between $\mathcal{JI}$ and $\mathcal{HDI}$, and between $\mathcal{ACC}$ and $\mathcal{CN}$. Also, from the distribution of the feature in the diagonals of Figure~\ref{fig:RG_FDM}, panels~\ref{fig:RG_FDM_b},~\ref{fig:RG_FDM_d},~\ref{fig:RG_FDM_f},~\ref{fig:RG_FDM_h},~\ref{fig:RG_FDM_j}, and~\ref{fig:RG_FDM_l}, it is obvious that the distribution of these features is not Gaussian. Most features of all $FDMs$ show low variance, except for the FDM of Facebook in Figures~\ref{fig:RG_FDM_i} and \ref{fig:RG_FDM_j}. This will affect the performance of the classifiers as we will see later.

Figure~\ref{fig:LF_FDM} shows the same analysis as presented in Figure~\ref{fig:RG_FDM} but for the \textit{Law Firm} dataset.
However, there are some differences in the properties of the features of the $FDM$s of the Law Firm networks. For example, the networks' $FDM$s have more variance for all features of the $FDMs$ of all networks. Also, the features are more correlated with each other when compared to the Research Group dataset.

\subsection{Classification results}
\label{subsec:classificationResults}
Next, we will present the results of the method proposed in this work. The results were first obtained for random graphs as a null model. The results of the null model were insignificant compared to the results presented here~\cite{Abufouda2015}.
 The results presented in this sections are based on the $\mathcal{SVM}$ classifier~\cite{cortes1995} with a Gaussian kernel. Table~\ref{tab:researchGroupResult} shows the results of the assessment for the Research Group dataset. The assessment results are satisfactory in terms of the evaluation metrics. The lower bound for the classification is $0.824$ in terms of the F-measure, which is good considering the very small data the $FDM$ of the network Coauthor contains (cf. Figure~\ref{fig:g2} to see how small this network is). Note that this lower bound was improved compared to our previous work in~\cite{abufouda2014} due to incorporating additional edge proximity measures. Surprisingly, the results in the table also show that the aggregated $FDM$ does not provide any noticeable advantage over the single networks. Having said that, the Lunch and the Work networks provided the best results, which suggests the friendship in the Research Group dataset is highly correlated with the Lunch network, which seems reasonable as we tend to have lunch often with our friends, but it is not necessarily that we coauthor with a friend.

\begin{table}[htbp!]
  \centering
  \begin{small}
    \caption{The prediction results for the Research Group dataset. Note that $k$-fold cross-validation was used when training and testing on $SN$.}
\begin{tabular}{|c|l|c|c|c|c|c|}
\hline
\multirow{2}{*}{Dataset} & \multicolumn{1}{c|}{\multirow{2}{*}{Train on}} & \multirow{2}{*}{Test on}      & \multicolumn{4}{c|}{Performance}   \\ \cline{4-7} 
                         & \multicolumn{1}{c|}{}                          &                               & $\mathcal{ACC}$     & $\mathcal{P}$     & $\mathcal{R}$ & $\mathcal{F}$ \\ \hline
\multirow{6}{*}{RG}      & $G_1$: Work                                       & \multirow{6}{*}{SN: Facebook} & 0.841 & 0.842 & 0.841  & 0.841     \\ \cline{2-2} \cline{4-7} 
                         &$G_2$: Coauthor                                   &                               & 0.822 & 0.827 & 0.822  & 0.824     \\ \cline{2-2} \cline{4-7} 
                         & $G_3$: Lunch                                      &                               & 0.843 & 0.835 & 0.843  & 0.839     \\ \cline{2-2} \cline{4-7} 
                         & $G_4$: Leisure                                    &                               & 0.837 & 0.835 & 0.836  & 0.836     \\ \cline{2-2} \cline{4-7} 
                         & Aggregated                                     &                               & 0.834 & 0.834 & 0.830  & 0.836     \\ \cline{2-2} \cline{4-7} 
                         & SN: Facebook                                       &                               & 0.833 & 0.829 & 0.830  & 0.832     \\ \hline
\end{tabular}
\label{tab:researchGroupResult}%
    \end{small}
\end{table}%

Table~\ref{tab:lawFirmResult} shows the classification results for the Law Firm dataset. The results in the table shows better performance compared to the Research Group dataset with a lower bound of $0.88$ for the F-measure.

\begin{table}[htbp]
  \centering
    \caption{The prediction results for the Law Firm dataset. Note that $k$-fold cross-validation was used when training and testing on $SN$.}
\begin{tabular}{|c|l|c|c|c|c|c|}
\hline
\multirow{2}{*}{Dataset} & \multicolumn{1}{c|}{\multirow{2}{*}{Train on}} & \multirow{2}{*}{Test on}    & \multicolumn{4}{c|}{Performance}   \\ \cline{4-7} 
                         & \multicolumn{1}{c|}{}                          &                             & $\mathcal{ACC}$     & $\mathcal{P}$     & $\mathcal{R}$ & $\mathcal{F}$ \\ \hline
\multirow{4}{*}{LF}      & G1: Cowork                                     & \multirow{4}{*}{SN: Friend} & 0.889 & 0.884 & 0.889  & 0.886     \\ \cline{2-2} \cline{4-7} 
                         & G2: Advice                                     &                             & 0.893 & 0.887 & 0.893  & 0.889     \\ \cline{2-2} \cline{4-7} 
                         & Aggregated                                     &                             & 0.885 & 0.879 & 0.885  & 0.881     \\ \cline{2-2} \cline{4-7} 
                         & SN: Friend                                     &                             & 0.972 & 0.984 & 0.919  & 0.947     \\ \hline
\end{tabular}
  \label{tab:lawFirmResult}%
\end{table}%

We think that the slight advantage in the performance of the Law Firm dataset over the Research Group dataset is due to the higher variance in the $FDM$'s features of the Law Firm as shown in Figure~\ref{fig:LF_FDM}. Even though the RG dataset has more networks, its aggregated $FDM$ did not show better results than the single networks. This indicates that for a better classification of the links, we need more features that capture the structure of the network, other than the edge proximity features that we used. In addition, the results indicate that directed networks may contain more patterns regarding the interaction among the members of these networks. 

\subsection{Comparing different classifiers}
\label{subsec:comparingDifferentClassifiers}
There are dozens of machine learning classifiers, and each has its advantages, limitations, and parameters to tune, which makes the selection of the appropriate classifier a difficult task. Table~\ref{tab:multipleClassifiers} shows a comparison of the performance of different classifiers. Based on the results in the table, the presented method showed close performance for most classifiers. Once again, the results of the LF dataset are slightly better than those of the RG dataset for all of the compared classifiers.

\begin{table}[htbp]
  \centering
  \caption{Comparison of the performance of different classifiers for the aggregated versions of the RG and the LF datasets. The compared classifiers are: The $\mathcal{KN}$: k-Nearest Neighbors vote~\cite{altman1992}; $\mathcal{SVM}$: Support Vector Machines~\cite{cortes1995};  $\mathcal{DT}$: decision trees~\cite{quinlan1986}; $\mathcal{NB}$: Naive Bayes~\cite{zhang2004}; $\mathcal{LR}$: Logistic Regression~\cite{walker1967}. We used the \textit{scikit-learn} Python package~\cite{scikitlearn2011}. } 
\begin{tabular}{|c|c|c|c|c|c|}
\hline
\multirow{2}{*}{Dataset} & \multirow{2}{*}{Classifier} & \multicolumn{4}{c|}{Performance}                                \\ \cline{3-6} 
                         &                             & $\mathcal{ACC}$ & $\mathcal{P}$ & $\mathcal{R}$ & $\mathcal{F}$ \\ \hline
\multirow{5}{*}{RG}      & $\mathcal{KN}$                          & 0.800           & 0.795         & 0.800         & 0.765         \\ \cline{2-6} 
                         & $\mathcal{SVM}$                         & 0.821           & 0.833         & 0.821         & 0.825         \\ \cline{2-6} 
                         & $\mathcal{DT}$                           & 0.800           & 0.806         & 0.800         & 0.804         \\ \cline{2-6} 
                         & $\mathcal{NB}$                          & 0.778           & 0.821         & 0.778         & 0.780         \\ \cline{2-6} 
                         & $\mathcal{LR}$                          & 0.827           & 0.825         & 0.827         & 0.827         \\ \hline\hline
\multirow{5}{*}{LF}      & $\mathcal{KN}$                         & 0.843           & 0.823         & 0.843         & 0.794         \\ \cline{2-6} 
                         & $\mathcal{SVM}$                          & 0.816           & 0.858         & 0.816         & 0.830         \\ \cline{2-6} 
                         & $\mathcal{DT}$                          & 0.880           & 0.870         & 0.880         & 0.870         \\ \cline{2-6} 
                         & $\mathcal{NB}$                          & 0.883           & 0.875         & 0.883         & 0.877         \\ \cline{2-6} 
                         & $\mathcal{LR}$                          & 0.868           & 0.878         & 0.868         & 0.872         \\ \hline
\end{tabular}
  \label{tab:multipleClassifiers}%
\end{table}%

\begin{figure}[htbp]
\center
  \noindent\makebox[\textwidth]{
\includegraphics[scale=0.4]{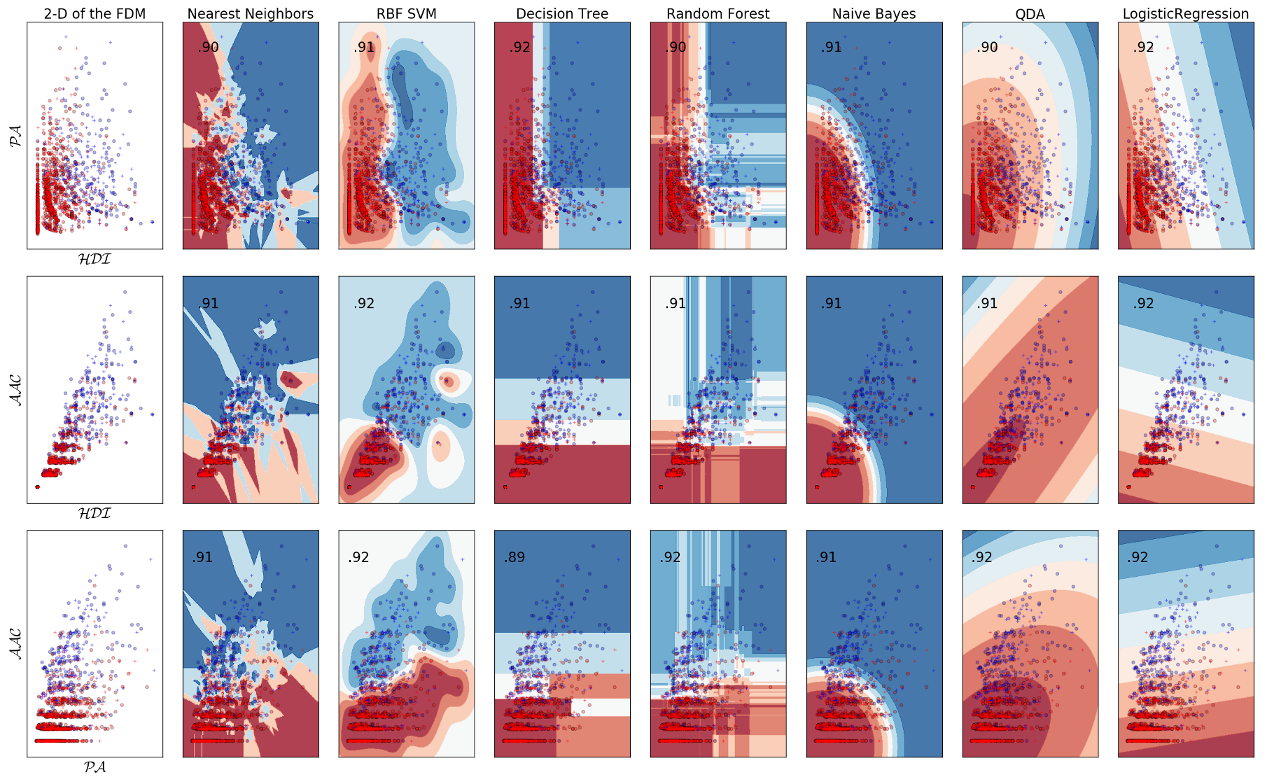}}
\caption {(Color online) The decision boundaries for different probability-based classifiers. We used 2-d scatter of the $FDM$ constructed from the aggregated networks of the RG dataset as an illustration. The leftmost panels are the 2-d features before the classification was performed. The red points are False instances, the blue points are True instances. The red "+" markers and the blue "+" markers are the False and the True instances to be classified by the classifier, i.e., the test samples. The other points, none "+" points, are the training points, where the training and the testing points were randomly selected with ratio 75:25, respectively. The other panels represent the classification results with the decision boundaries. The number in the top-left is the accuracy of the classification, and the gradient of the colored areas represents the probability. For example, the darker the blue area, the higher the probability that the points in this area are true. The classifiers used are those classifiers that give a probability as a classification result, and they are, namely: $\mathcal{KN}$: k-Nearest Neighbors vote~\cite{altman1992}; $\mathcal{SVM}$ with Gaussian kernel~\cite{cortes1995}; $\mathcal{DT}$: decision trees~\cite{quinlan1986}; Random Forests~\cite{ho1995random}; $\mathcal{NB}$: Naive Bayes~\cite{zhang2004}; $\mathcal{QDA}$: the Quadratic Discriminant Analysis~\cite{cover1965geometrical}; $\mathcal{LR}$: Logistic Regression~\cite{walker1967}. }
\label{fig:decision_boundaries_rg}
\end{figure}

Another aspect that is important when talking about different classifiers is the resulting decision boundaries and how good they are.
Figure~\ref{fig:decision_boundaries_rg} shows the decision boundaries for different classifiers. The figure shows that linear models, like linear $\mathcal{DT}$ and $\mathcal{LR}$, were not able to really discriminate between the False and the True instances efficiently. Additionally, the figure shows that the accuracy metric is a useless measure as it is not informative for the case of the $FDM$, whose labels are highly imbalanced. For example, let us take a closer look at the $\mathcal{QDA}$ classifier for the second panel, the attributes $\mathcal{HDI}$ vs $\mathcal{AAC}$. The accuracy of the classifier is $0.91$ which is considered high. Having said that, the panel shows that \textit{all} of the points were classified in the red area, which ignores the True instances and make it hard to find a binary threshold to produce binary results. Such a behavior indicates that the accuracy is not a good measure to use if we have imbalanced data. On the other hand, classifiers that use kernels (a method to transferring the non-linearly separable data into linearly separable data by transforming the data into a higher dimension) showed good discrimination between the False and the True instances. An example of this is $\mathcal{SVM}$ with Gaussian kernel~\cite{cortes1995}, the third column in Figure~\ref{fig:decision_boundaries_rg}. From the figure, it is clear that the $\mathcal{SVM}$ with Gaussian kernel is able to find disjoint areas for the data points, which helps in producing good classification results.

\begin{figure}[!ht]
    \centering
    \begin{subfigure}[b]{0.3\textwidth}
    	\centering
        \includegraphics[width=\linewidth]{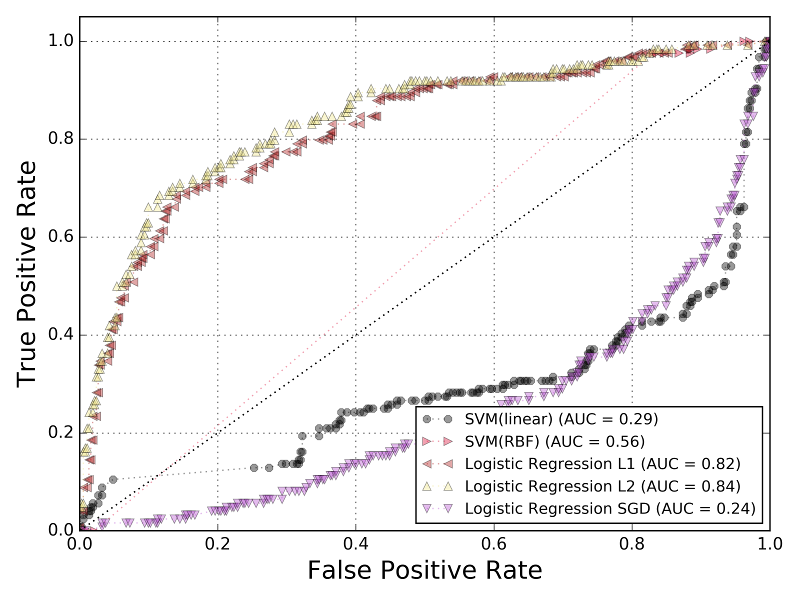}
        \caption{Coauthor}
        \label{fig:AUC_RG_LF_a}
    \end{subfigure}\hspace{1em}
    \begin{subfigure}[b]{0.3\textwidth}
        	\centering
        \includegraphics[width=\linewidth]{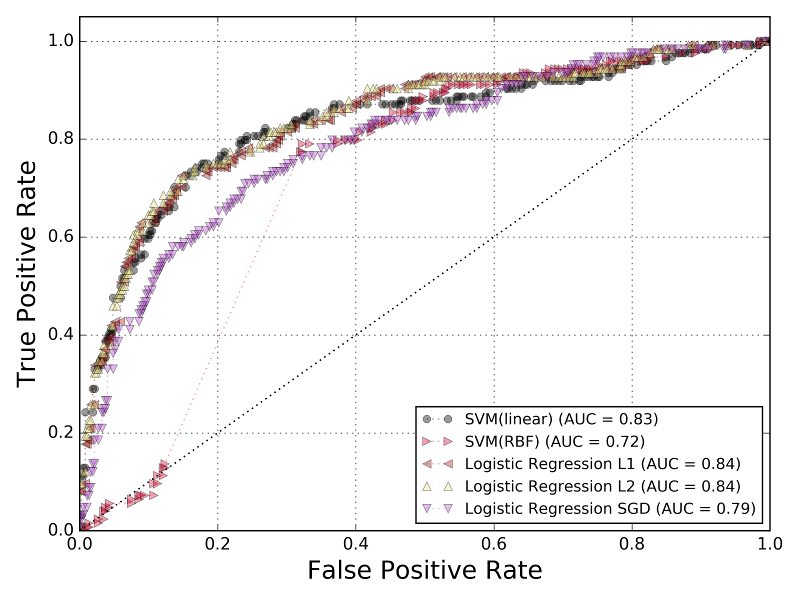}
        \caption{Leisure}
        \label{fig:AUC_RG_LF_b}
    \end{subfigure}
    \begin{subfigure}[b]{0.3\textwidth}    
    	\centering
        \includegraphics[width=\linewidth]{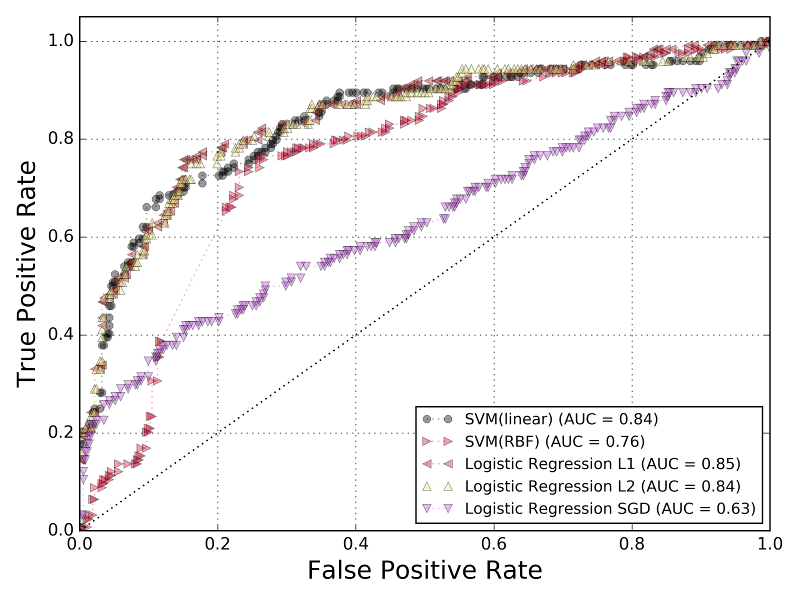}
        \caption{Lunch}
        \label{fig:AUC_RG_LF_c}
    \end{subfigure}
    \begin{subfigure}[b]{0.3\textwidth}
        	\centering
        \includegraphics[width=\linewidth]{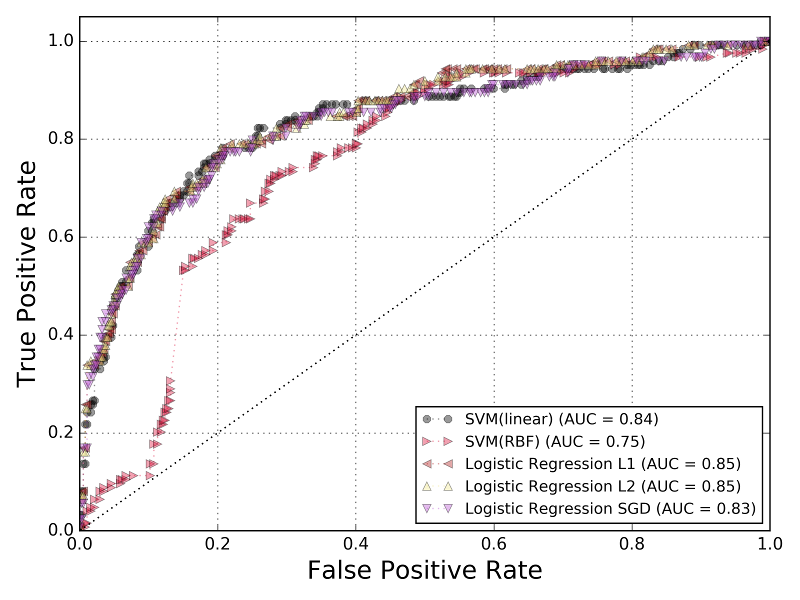}
        \caption{Work}
        \label{fig:AUC_RG_LF_d}
    \end{subfigure}
    \begin{subfigure}[b]{0.3\textwidth}
        	\centering
        \includegraphics[width=\linewidth]{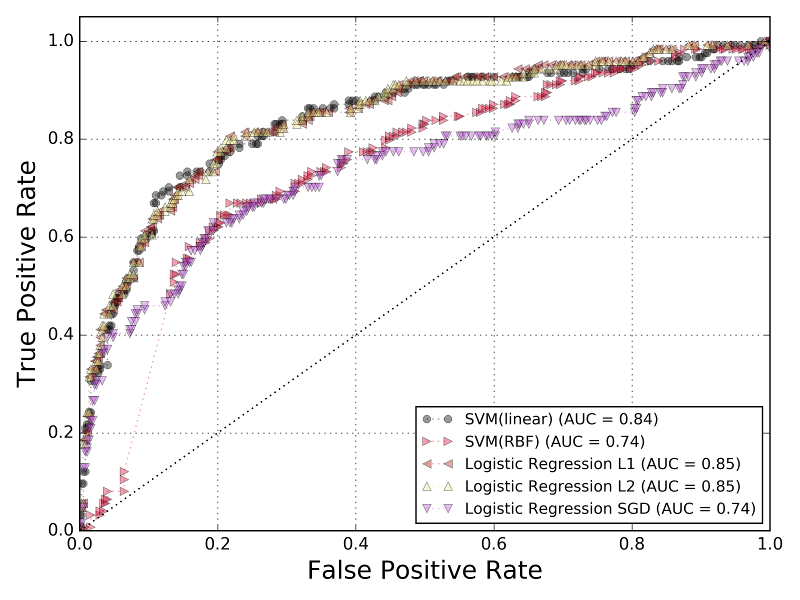}
        \caption{RG aggregated}
        \label{fig:AUC_RG_LF_e}
    \end{subfigure}
    \begin{subfigure}[b]{0.3\textwidth}
        	\centering
        \includegraphics[width=\linewidth]{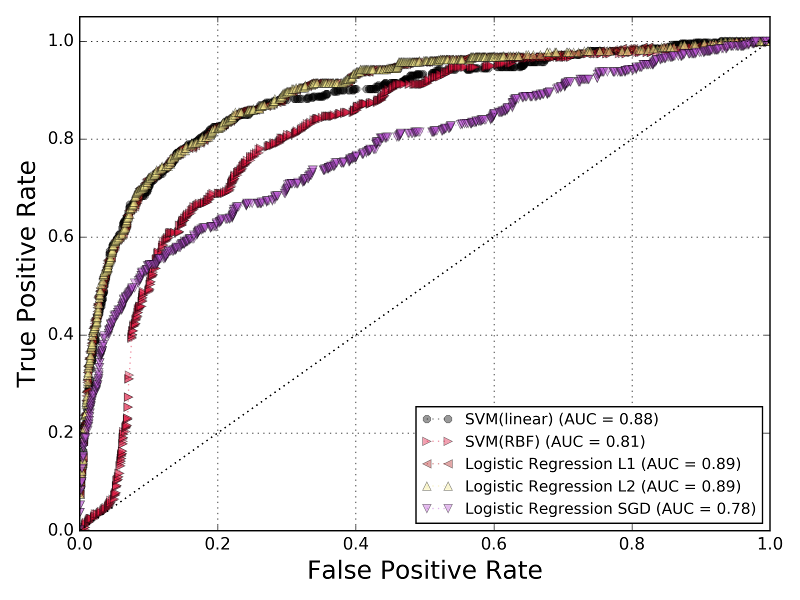}
        \caption{Advice}
        \label{fig:AUC_RG_LF_f}
    \end{subfigure}
        \begin{subfigure}[b]{0.3\textwidth}
            	\centering
        \includegraphics[width=\linewidth]{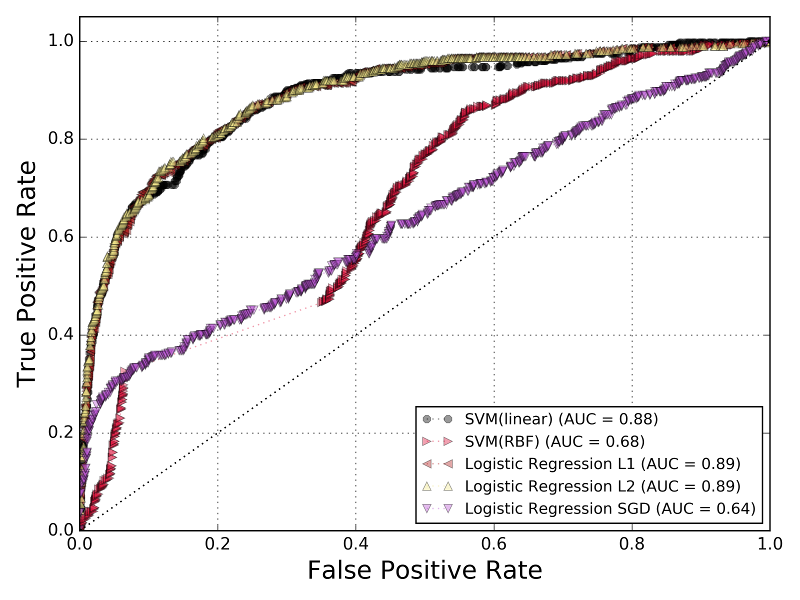}
        \caption{Coworker}
        \label{fig:AUC_RG_LF_g}
    \end{subfigure}
        \begin{subfigure}[b]{0.3\textwidth}
            	\centering
        \includegraphics[width=\linewidth]{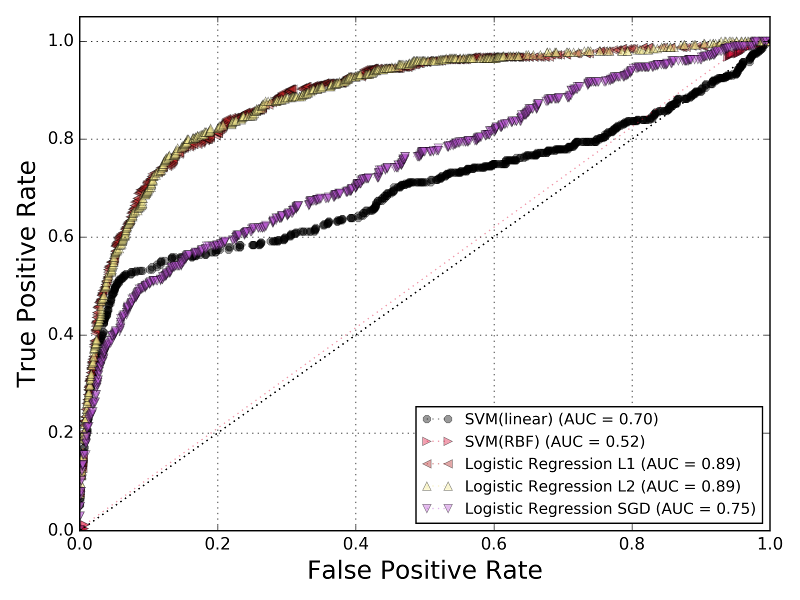}
        \caption{LF aggregated}
        \label{fig:AUC_RG_LF_h}
    \end{subfigure}
    \caption{(Color online) The area under the ROC curve for $\mathcal{SVM}$ with linear and Gaussian kernels and for $\mathcal{LR}$ with $L1$ and $L2$ regularization and with a stochastic gradient descent optimization algorithm.}
     \label{fig:AUC_RG_LF}
\end{figure}

Another way to compare the performance of different classifiers is to use the area under the ROC curve. Figure~\ref{fig:AUC_RG_LF} shows the AUC for $\mathcal{SVM}$ and $\mathcal{LR}$ with different tuning parameters. Figure~\ref{fig:AUC_RG_LF_a} again shows that the linear models are not robust and are not able to provide a good classification. Figures~\ref{fig:AUC_RG_LF}, panels~\ref{fig:AUC_RG_LF_b},~\ref{fig:AUC_RG_LF_c},~\ref{fig:AUC_RG_LF_d}, ~\ref{fig:AUC_RG_LF_e},~\ref{fig:AUC_RG_LF_f},~\ref{fig:AUC_RG_LF_g}, and~\ref{fig:AUC_RG_LF_h} show that $\mathcal{SVM}$ with Gaussian kernel provided a stable performance, which is why we used it for the results presented in Section~\ref{subsec:classificationResults}.

\subsection{From binary classification to tie strength ranking}
\label{subsec:fromBinaryToRanking}
\begin{figure}[!ht]
    \centering
    \begin{subfigure}[b]{0.3\textwidth}
    	\centering
        \includegraphics[width=\linewidth]{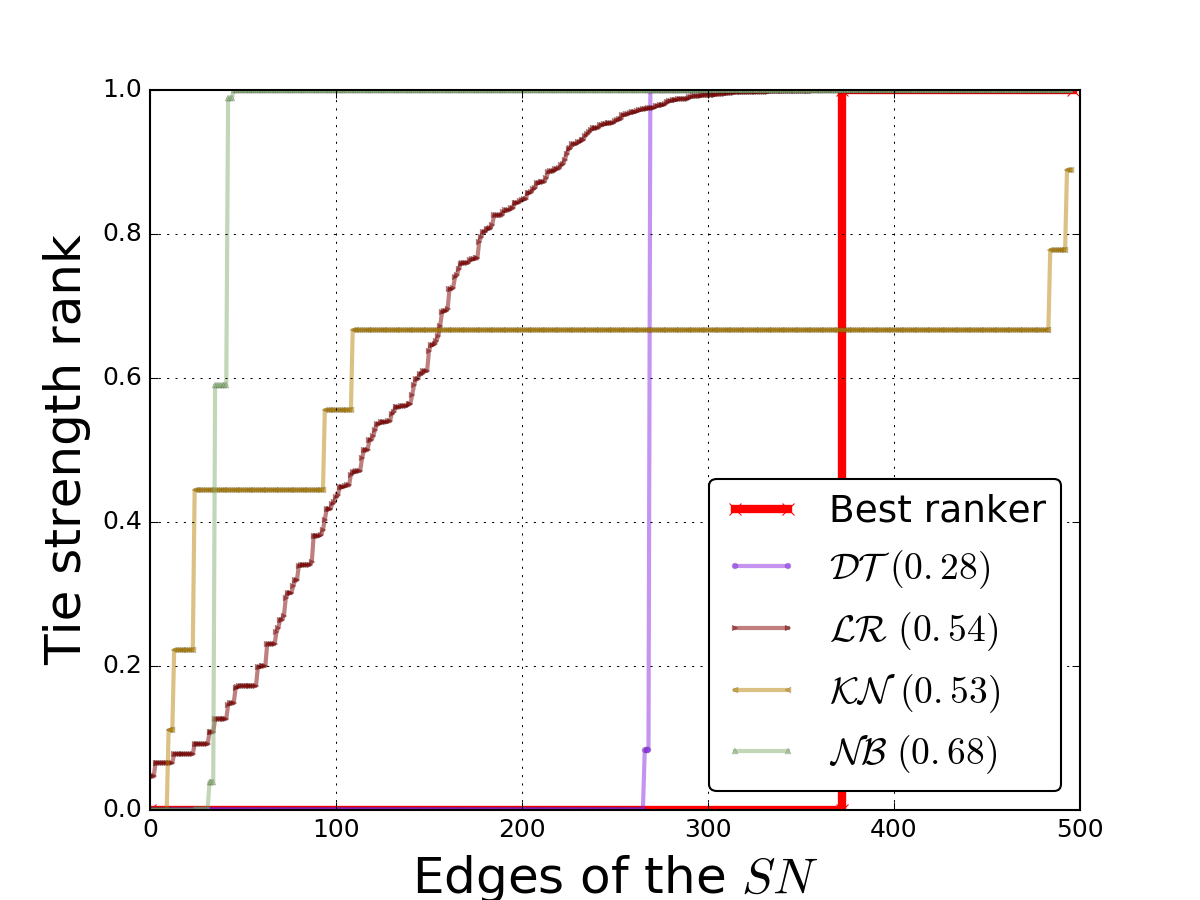}
        \caption{Coauthor}
        \label{fig:Coauthor}
    \end{subfigure}\hspace{1em}
    \begin{subfigure}[b]{0.3\textwidth}
        	\centering
        \includegraphics[width=\linewidth]{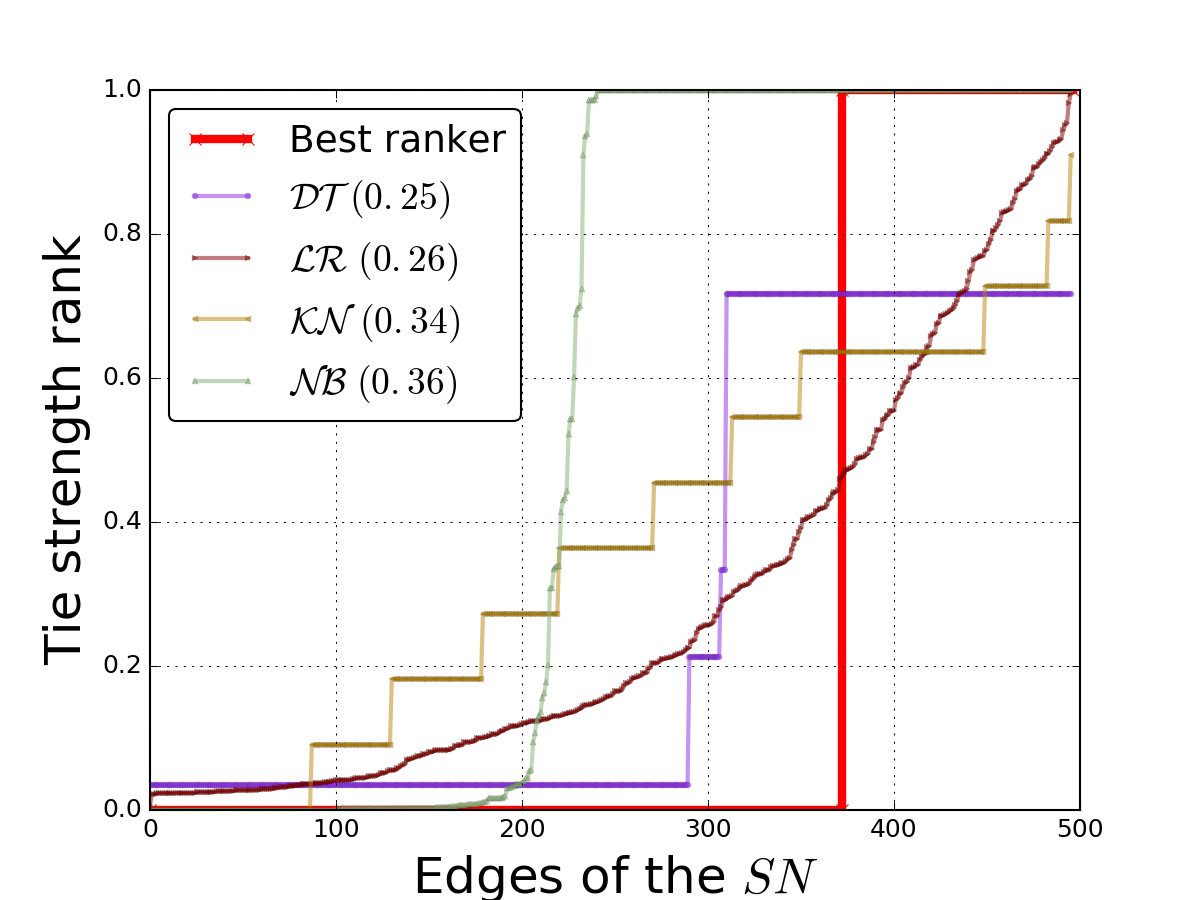}
        \caption{Leisure}
        \label{fig:Leisure}
    \end{subfigure}
    \begin{subfigure}[b]{0.3\textwidth}    
    	\centering
        \includegraphics[width=\linewidth]{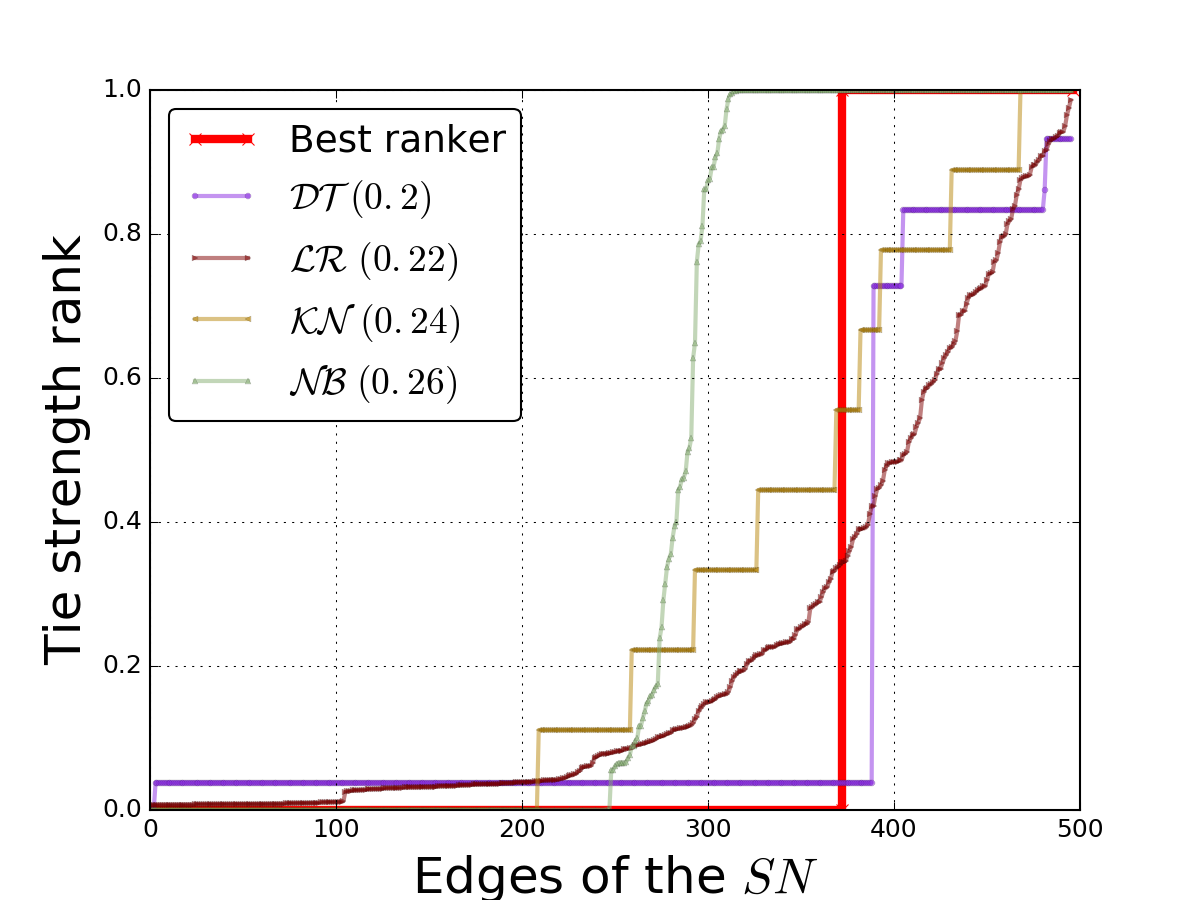}
        \caption{Lunch}
        \label{fig:Lunch}
    \end{subfigure}
    \begin{subfigure}[b]{0.3\textwidth}
        	\centering
        \includegraphics[width=\linewidth]{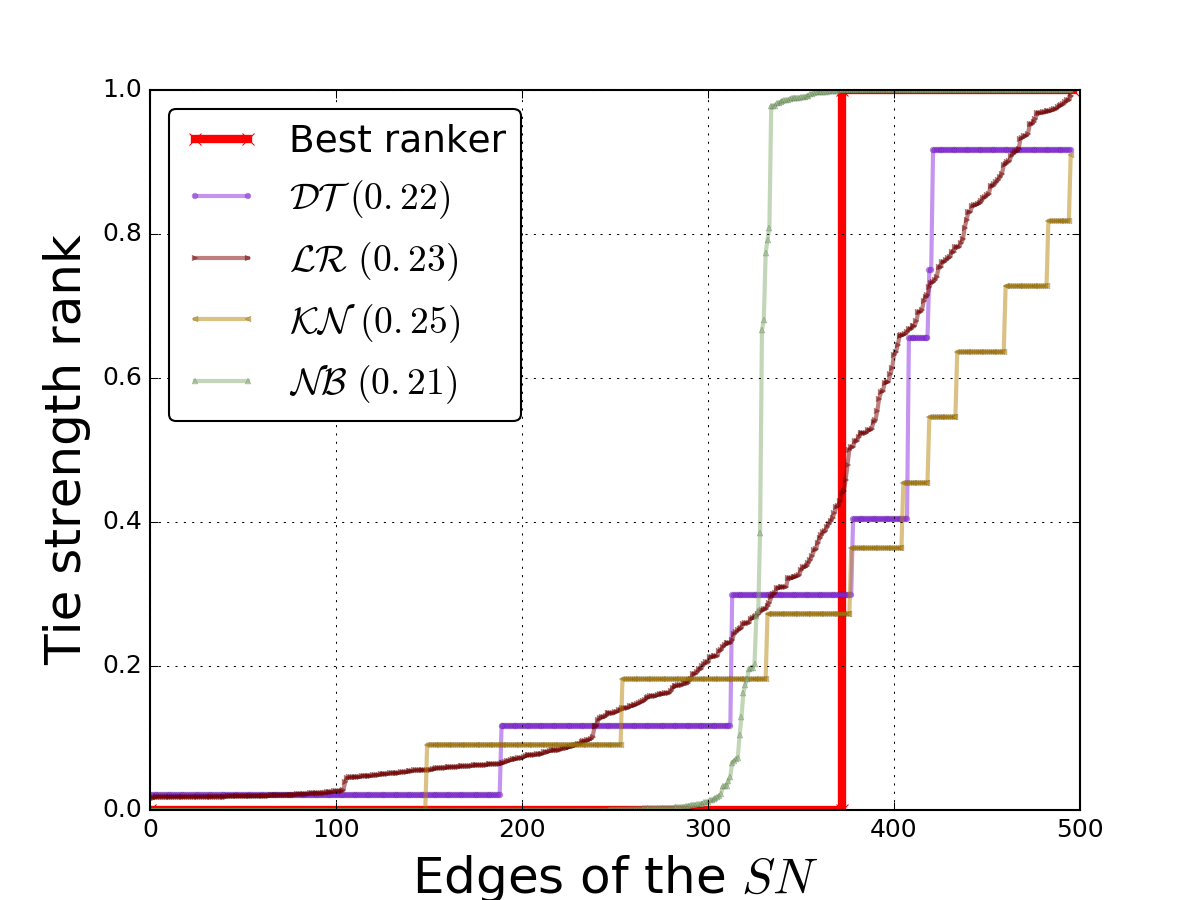}
        \caption{Work}
        \label{fig:work}
    \end{subfigure}
    \begin{subfigure}[b]{0.3\textwidth}
        	\centering
        \includegraphics[width=\linewidth]{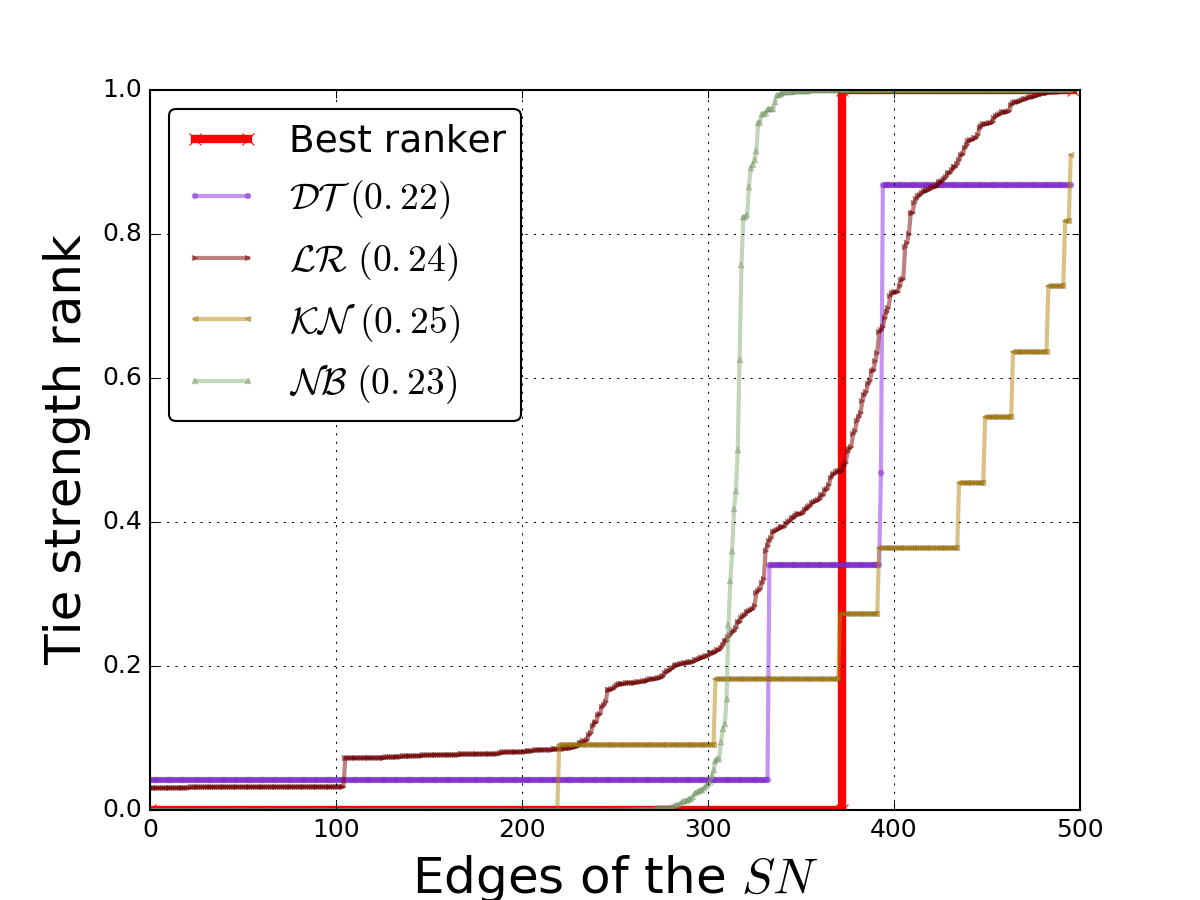}
        \caption{RG aggregated}
        \label{fig:aggregated_rg}
    \end{subfigure}
    \begin{subfigure}[b]{0.3\textwidth}
        	\centering
        \includegraphics[width=\linewidth]{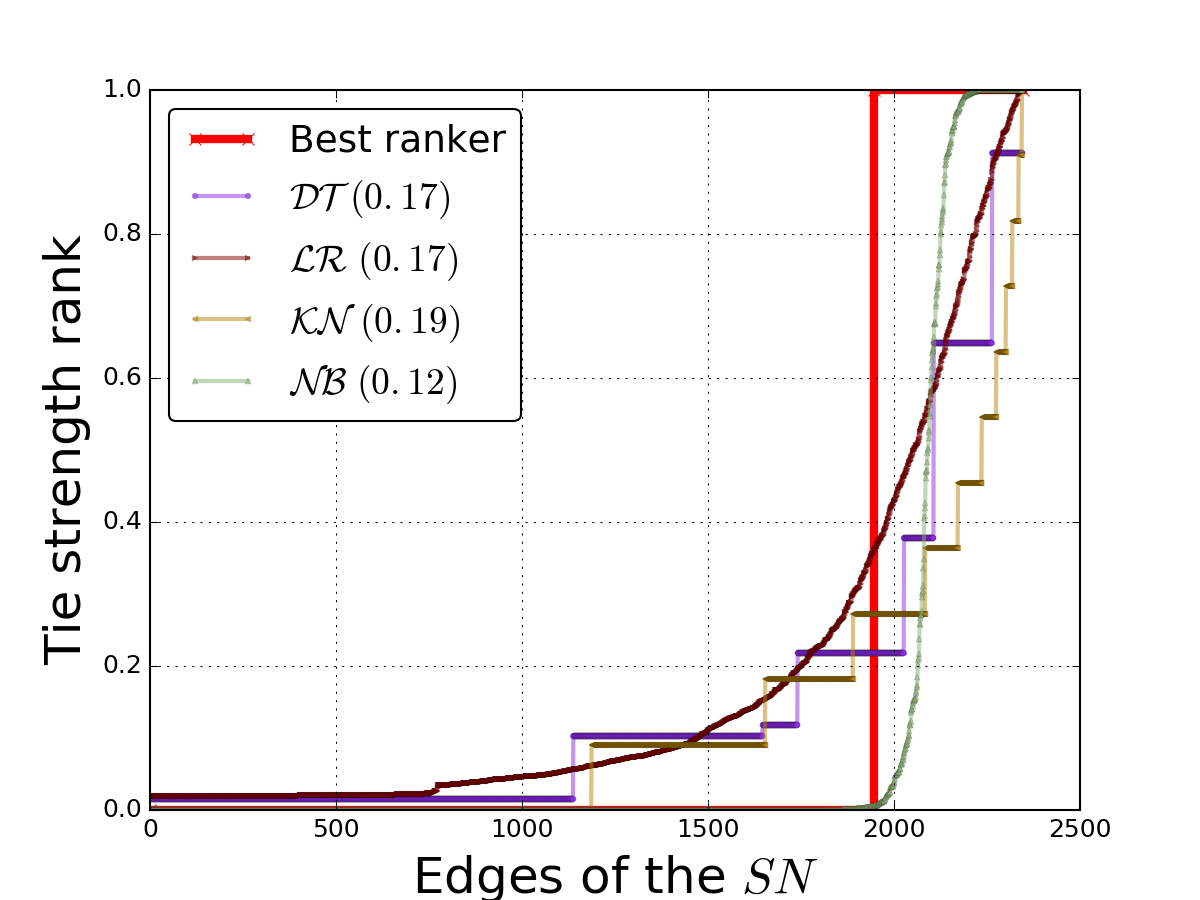}
        \caption{Advice}
        \label{fig:advice}
    \end{subfigure}
        \begin{subfigure}[b]{0.3\textwidth}
            	\centering
        \includegraphics[width=\linewidth]{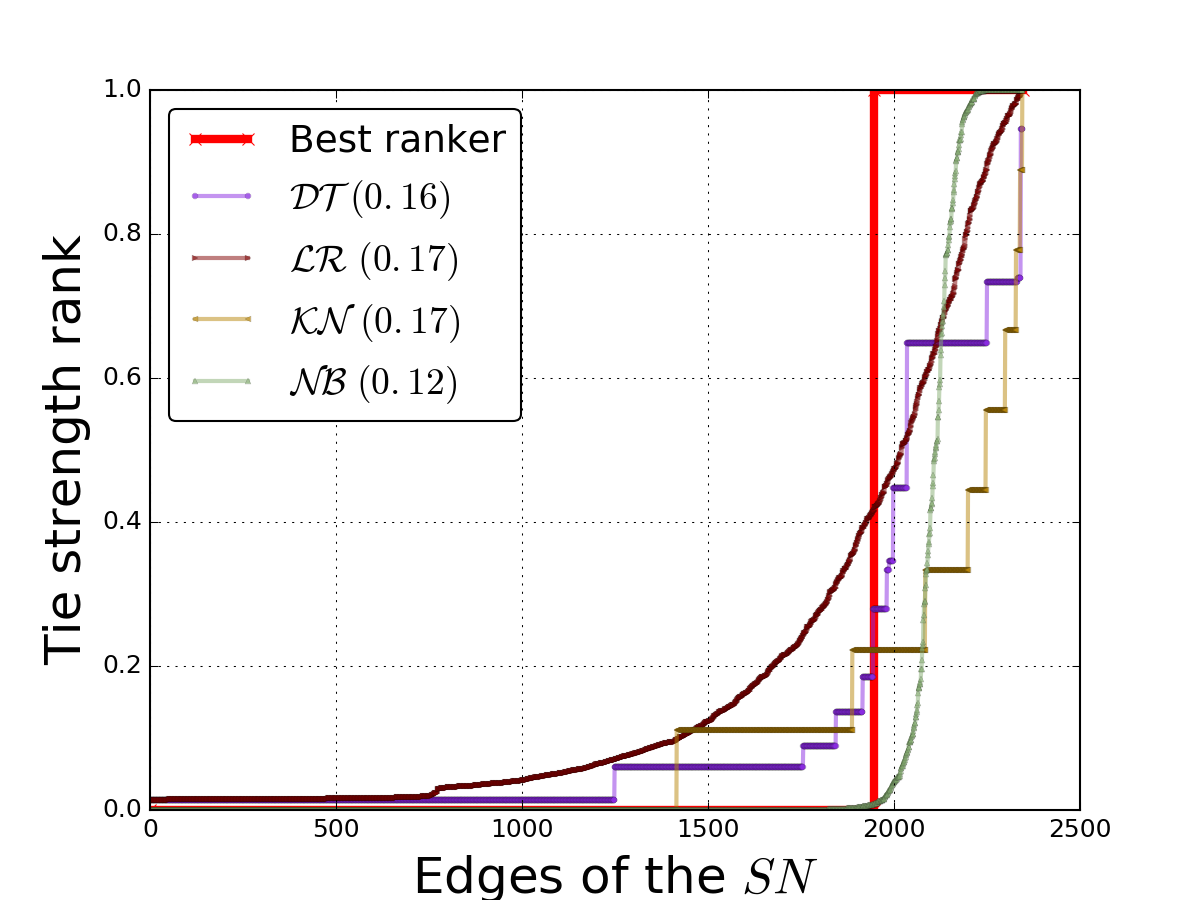}
        \caption{Coworker}
        \label{fig:coworker}
    \end{subfigure}
        \begin{subfigure}[b]{0.3\textwidth}
            	\centering
        \includegraphics[width=\linewidth]{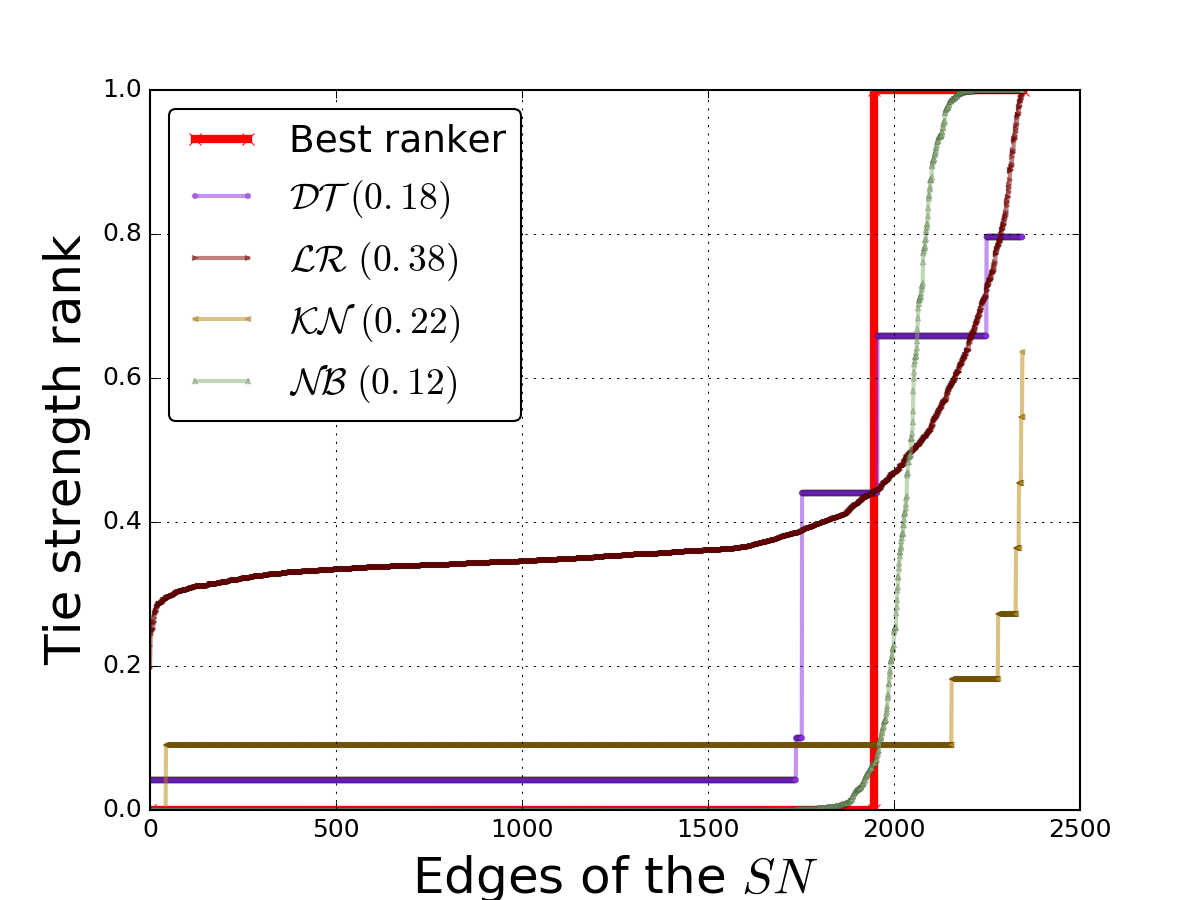}
        \caption{LF aggregated}
        \label{fig:aggregated_lf}
    \end{subfigure}
    \caption{ (Color online) The Tie strength ranking for the social network using the associated networks. The x-axis represents the edges in the $SN$ ranked by their strength according to the ranking results; the y-axis is the tie strength rank. The best ranker, in bold red, is simply the step function on the number of edges in the social network. The best ranker is used to compare the goodness of the ranking using the proposed method. In the legend, different classifiers are used to estimate the probabilities. The numbers beside the names of the classifiers are the errors in the ranking. This error is calculated as: $\sum_{e\in E_{SN}}{|e_{obs} -e_{real}|}$, where $E_{SN}$ is the set of edges being ranked in the $SN$, $e_{obs} \in [0,1] $ is the probability of having an edge $e$ in the $SN$, and $e_{real} \in \{1,0\}$ which means whether $e$ is a true-positive or a true-negative, respectively. }
     \label{fig:Ranking_RG_LF}
\end{figure}

In some scenarios, the links of a social network need to be ranked by the tie strength between the members. The proposed method can also give a continuous range of value between $0$ and $1$, instead of having two classes, using probabilistic classifiers: classifiers that produce a probability value instead of a binary class then finding a threshold for binarizing the resulted probabilities. These probabilities are used here as a\textit{ tie strength rank} of the edges in the social network being assessed.
Figure~\ref{fig:Ranking_RG_LF} shows the ranking results of the $SN$ in the RG and the LF datasets using different classifiers.
Our assumption here is that the \textit{best} ranking for the edges of the $SN$ is a step function that changes the values of the edge from \textit{zero} to \textit{one} on the number of true-negative edges in the ground truth network. That means, for undirected $SN$ with $n$ nodes and $m$ edges we have $m$ edges with tie strength $one$ and $ {n \choose 2} -m $ edges with tie strength $zero$. Then, the predicted tie strength for all edges, including the true-negatives in the ground truth, is compared to the best ranking using the following error measure:  $\sum_{e\in E_{SN}}{|e_{obs} -e_{real}|}$, where $E_{SN}$ is the set of edges being ranked, including true-negatives, in the $SN$, $e_{obs} \in [0,1] $ is the probability of having an edge $e$ in the $SN$, and $e_{real} \in \{1,0\}$ which means whether $e$ is a true-positive or a true-negative, respectively.
The closer the results to the step function, the better the ranking (cf. Figure~\ref{fig:Ranking_RG_LF}) . 

Figure~\ref{fig:Ranking_RG_LF} shows the results of the ranking using the proposed method. $\mathcal{NB}$ and $\mathcal{KN}$ provided the best ranking among all of the classifiers we used. For the RG dataset, the best link ranking, in terms of error ranking as explained earlier, was achieved using the Lunch network with error $20\%$ and by the Work network with error $21\%$ using the $\mathcal{DT}$ and $\mathcal{NB}$, respectively. For the LF dataset, the best ranking was achieved using the any of the networks in the dataset with $12\%$ error using the $\mathcal{NB}$. As in the assessment results presented in the previous section, the ranking results of the $SN$ of the LF is better than the ranking of the RG's $SN$. Again, we think that the directed networks embrace more information about the structure and the relationships among their members.\\

\subsection{Noisy-edges identification}
\label{subsec:noiseidentification}
The results in Section~\ref{subsec:classificationResults} showed how the proposed method is only good in finding true-positive and true-negative edges in a social network. The datasets used in this work do not contain any noisy-edges in the social network, which does not allow a proper validation for noise (false-positives) identification in their original form. Thus, we injected noisy-edges in the $SN$ and tested the method to find how good it is in finding them. To do so, we added $k$ edges to the $SN$ such that $k= \floor*{({n \choose 2} -m) \times r} $ edges, where $m$ is the number of edges in the $SN$, i.e., the true-positives, and $r$ is the percentage of edges to be added. The resulted network is called $SN_{disguised}$.
For example, if $r=1$ then the $SN_{disguised}$ is a complete network. Then, we predicted only these $k$ edges using the method. The \textit{success rate} is defined as the number of edges which were predicted as false-positive divided by $k$. We used different values for $r$ ranging between $0.1$ and $1.0$. The results of the noise identification came as follows. For the research group dataset, the success rates were  $0.34, 0.94, 0.94, 0.95,$ and $0.97$ when training on Coauthor, Leisure, Lunch, Work, and the aggregated version, respectively, and testing only the $k$ edges in $SN_{disguised}$.
As the edges to be added to the $SN$ were randomly selected, the results were obtained as the mean of $10$ runs for each $r$ averaged by the number of values used of $r$. The poor performance for the coauthor network is because it is very small network, and it hardly captures a good structure for the relationships among the members. For the law firm dataset, the success rate was $0.99$, for all networks with the same settings of $r$ as in the research group dataset. The noise identification in the law firm dataset was higher than the research group dataset. However, the success rate for the aggregated networks of the research group showed a better performance than the performance of the best network, the work network.

\section{Discussion}
\label{sec:discussion}
The proposed method showed a good potential in both link classification and tie strength ranking. It seems that machine learning can effectively be used for the network based features. In this section we provide our final thoughts about the problem addressed in the paper, the used method, and the limitations. 

\subsection{The importance of link assessment and tie ranking}
Addressing the link assessment problem is crucial in today's life, where online social media contain a lot of spam, ads-intensive websites, and inaccurate news. We strongly think that identifying noisy links in social networks contributes in eliminating these problems and reducing their impacts. Tie strength ranking, on the other hand, can also improve the quality of information spread in online social networks. For example, automatic ranking of the friends list on Facebook might led to better news feed, friends recommendations, and better targeted ads, just to name a few. Thus, the work presented in this paper has actionable insights on online social networks.\\
It seems that the tie strength problem explicitly includes link assessment. However, the two problems should be separately handled because the cost of link assessment may be lower than the cost of tie strength ranking. One reason for that is the hardness of getting the ground truth of real tie strengths between the nodes of a network. This reason pushed some researchers to concentrate only on the strong ties, like the work presented in the related work section. Another reason is that, the link classification is sufficient in certain applications such as spam detection. Moreover, link classification can be a preprocessing step for many network based analysis tasks, such as community detection, where eliminating noise edges may provide more meaningful communities.
Thus, we emphasis the distinction between the two problems.
\subsection{Classification methods for network-based features}
\textbf{Features correlation:} The major network-based features for link proximity are based on common neighbors $\mathcal{CN}$, which makes most of the features highly correlated to each other. Having said that, highly correlated features might be a problem in classification, especially with small number of training data. Thus, devising new link proximity measures that are not based on the number of common neighbors is important.\\
\textbf{Data imbalance:} Imbalanced dataset is a common challenge in classification problems. In social networks case, this problem is more vivid because most social networks are sparse. Thus, any edge proximity-based model is inherently imbalanced. Many techniques exists in the literature to avoid the classification limitation under imbalanced datasets \cite{kotsiantis2006}. In this work, we used SMOTE (Synthetic Minority Over-sampling Technique) \cite{chawla2002}, which did not give any improvement in the prediction performance due to the small datasets we have. The literature contains a lot of techniques that can be used with larger datasets to handle the imbalanced nature of some datasets \cite{kotsiantis2006}. \\
\textbf{Classifier selection:} The decision boundaries are helpful to select a good classifier for the used dataset. Learning and optimization processes are computationally expensive, and experimenting different classifiers with different parameters is always laborious task. Thus, experimenting on sample of the data to select the best classifier is crucial. To handle that, decision boundaries, like what presented in Figure \ref{fig:decision_boundaries_rg}, helps a lot in understanding the data that we have and also to select the best classifier for subsequent optimization.
Linear classifiers showed poor performance as the constructed $FDM$ is not linearly separable. Thus, using classifiers with kernels showed better performance. Additionally, $\mathcal{KN}$ and $\mathcal{DT}$ showed promising results for the ranking problem. It turned out that these two classifiers provided good probabilities for approximating the tie strength in the $SN$, but bad thresholds for the binary classification.\\
Data cleansing, especially outliers removal, may improve the prediction results. However, in this work we did not remove any data as the used datasets are relatively small. For example, when removing all data that is 3 times the standard deviation away from the mean, the results were not as good as the presented in the results section, though, removing outliers might improve the results in the case of having larger datasets.\\ 
\textbf{Baseline comparison:} To provide more confidence for the results, we conducted the experiments on random graphs as a null model. The results of the random graphs were uncomparable to the results of the real used datasets\footnote{More details about the results of the random graphs can be found in our earlier work~\cite{Abufouda2015}.}. Additionally, we tested the model against a classifier that uses one simple rule as a baseline prediction. The results of the presented method using the classifiers presented in Section~\ref{sec:relatedwork} were significantly better than the baseline classifier. Finally, a random classifier was used as another baseline classifier, (cf. Figure~\ref{fig:AUC_RG_LF}). The results of the used classifiers were significantly better than random classifier. Thus, we strongly think that the results provided in this paper are significant and are not due to any random chances.
\subsection{Limitations}
The presented method used data from a social network itself in addition to external information. The external information may not always be available, which represents a challenge. Moreover, the existence of the ground truth data for the tie strength ranking is hard to attain. Thus, we resort to the binary ranker as a gold standard measure to evaluate the tie ranking provided by the method.

\section{Summary}
\label{sec:summary}
In this paper, we presented a method for link assessment and link ranking of the links of online social networks using external social interaction networks. The proposed method employed machine learning classification techniques to perform the link assessment via label classification based on edge-proximity measures. We have conducted experiments on two different datasets that contain a friendship social network in addition to the external social interactions. The link assessment results, in terms of the F1-score and the accuracy, were satisfactory compared to baseline predictors. The results show that it is possible to assess the links in a social network using external social interactions. Additionally, link ranking has also been performed using probabilistic binary classifiers.
The intensive study of the features used in this work and the conducted experiments revealed insights about using machine learning for network-based features. These insights are about (1) features correlation and its effect on the classification; (2) label imbalance handling; (3) goodness of the decision boundaries of the used classifiers; (4) classifier selection for both link assessment and link ranking.\\
From network perspective, the results of the used datasets suggest that directed networks embrace more building structures that enable better link assessment and link ranking compared to the undirected networks. Also, the results suggest that one external interaction networks embraces enough information to assess or rank the links in the social networks. It seems that for a set of persons, a social interaction outside the social network is enough to know much information about their social relationships. From machine learning perspective, the results achieved in this work, for both link assessment and link ranking, show that network-based features can be used in analyzing networks and building prediction models. Additionally, we discovered that some classifiers are good in providing a binary classification for link assessment, while some others are good in providing a probability range for link ranking.\\
Future work includes utilizing new features that are not based on common neighbors, implementing techniques for handling imbalanced labels, and incorporating feature selection before using the classifiers on the whole set of features.

\bibliographystyle{abbrv}
\bibliography{references}

\end{document}